# Benchmarking LLM-based agents for single-cell omics analysis

Yang Liu[1,2,10], Lu Zhou[1,2,10], Xiawei Du[7,8], Ruikun He[6*], Xuguang Zhang[3,9*], Rongbo Shen[1,2*], Yixue Li[1,2,3,4,5*]

1, Guangzhou National Laboratory, Guangzhou, 510005, China.
2, GMU-GIBH Joint School of Life Sciences, The Guangdong-Hong Kong-Macao Joint Laboratory for Cell Fate Regulation and Diseases, Guangzhou Medical University, Guangzhou, 511436, China.
3, Shanghai Institute of Nutrition and Health, Chinese Academy of Sciences, Shanghai, 200030, China.
4, Key Laboratory of Systems Health Science of Zhejiang Province, School of Life Science, Hangzhou Institute for Advanced Study, University of Chinese Academy of Sciences, Hangzhou, 310024, China.
5, School of Life Sciences and Biotechnology, Shanghai Jiao Tong University, Shanghai, 200240, China.
6, BYHEALTH Institute of Nutrition & Health, Guangzhou, 510663, China.
7, Department of Computer Science and Technology, Tsinghua University, Beijing, 100084, China
8, Dawning Information Industry (Beijing) Co.,Ltd, Beijing, 100176, China
9, Feihe Reseach Institute, Heilongjiang Feihe Dairy Co., Ltd., C-16,10A Jiuxianqiao Road, Chaoyang District, Beijing, 100015, China
10, These authors contributed to the work equally and should be regarded as co-first authors.

*Corresponding authors: Ruikun He (herk@by-health.com), Xuguang Zhang (xgzhang@sinh.ac.cn), Rongbo Shen (shen_rongbo@gzlab.ac.cn), Yixue Li (li_yixue@gzlab.ac.cn).

# Abstract

Background:

The surge in single-cell omics data exposes limitations in traditional, manually defined analysis workflows. AI agents offer a paradigm shift, enabling adaptive planning, executable code generation, traceable decisions, and real-time knowledge fusion. However, the lack of a comprehensive benchmark critically hinders progress.

Results:

We introduce a novel benchmarking evaluation system to rigorously assess agent capabilities in single-cell omics analysis. This system comprises: a unified platform compatible with diverse agent frameworks and LLMs; multidimensional metrics assessing cognitive program synthesis, collaboration, execution efficiency, bioinformatics knowledge integration, and task completion quality; and 50 diverse real-world single-cell omics analysis tasks spanning multi-omics, species, and sequencing technologies. Our evaluation reveals that Grok3-beta achieves state-of-the-art performance among tested agent frameworks. Multi-agent frameworks significantly enhance collaboration and execution efficiency over single-agent approaches through specialized role division. Attribution analyses of agent capabilities identify that high-quality code generation is crucial for task success, and self-reflection has the most significant overall impact, followed by retrieval-augmented generation (RAG) and planning.

Conclusions:

This work highlights persistent challenges in code generation, long-context handling, and context-aware knowledge retrieval, providing a critical empirical foundation and best practices for developing robust AI agents in computational biology.

# I. Background

Revolutionary breakthroughs in single-cell omics technologies—including single-cell transcriptomics, spatial transcriptomics, and integrated multi-omics profiling—are redefining the precision of biological research, ushering in an era of "cell-level resolution" [1]. Driven by international consortia like the Human Cell Atlas [2], public repositories currently house multimodal data for over 50 million single cells. The exponential growth in single-cell data complexity contrasts with the linear evolution of analytical methods, exposing limitations in traditional analysis paradigms. Specifically, traditional analysis heavily depends on manual pre-selection of algorithm combinations and parameter tuning, making the results non-objective and operator-dependent [3]. Critical steps, from feature selection to predictive inference, lack transparent decision-making pathways, bringing limited interpretability and hindering biological insight [4]. Traditional analysis also suffers from delayed knowledge fusion. Built-in reference databases within analysis tools often lag by more than 6 months and integrating knowledge across disparate sources requires substantial manual curation [5].

To address the challenges posed by the complexity (e.g., dynamic features, ultra-high dimensionality, and multimodal associations) of single-cell omics data [6] to traditional analysis paradigms and overcome the bottleneck in "data-algorithm-knowledge" optimization, AI agents emulate the human expert’s cognitive cycle of "hypothesis generation-experimental validation-iterative refinement", achieving a paradigm shift across three aspects. First, self-adaptive workflow planning and execution [7]. Agents interpret natural language instructions, comprehend task objectives, and autonomously plan optimal analytical workflows including algorithm selection and parameter configuration tailored to the specific data characteristics. They then generate and execute code to perform the analysis. Second, traceable, cross-disciplinary decision-making. Configured with specialized roles mirroring multidisciplinary experts, agents engage in semantic-level interactions [8]. This enables reasoning with transparent "thought chains" that trace logic from data inputs through analytical steps to biological conclusions, enhancing interpretability and trust. Third, real-time knowledge fusion. Agents integrate continuously with up-to-date biological databases and literatures. Leveraging Retrieval-Augmented Generation (RAG) technology [9], they dynamically retrieve and incorporate pertinent knowledge to validate analytical choices and ensure the biological plausibility of both workflows and results. Finally, AI agents adopt a "perception-reasoning-execution" architecture: The perception layer parses natural language, integrates prepared data, and connects to relevant knowledge bases; the reasoning layer then decomposes tasks, designs tailored analytical pipelines, and formulates the analysis plan; the execution layer generates executable code, runs analyses securely, and delivers interpretable results.

Despite the significant potential of AI agents in single-cell omics analysis, the field currently lacks a comprehensive, standardized benchmarking evaluation system. A major focus in many bio-oriented agent studies lies in proposing novel agent architectures and evaluating them on a limited set of specific tasks or use cases [10,11,12]. For instance, SpatialAgent targets three types of spatial omics applications, while Biomni is benchmarked on two QA-style reasoning benchmarks—Humanity's Last Exam [13] and Lab-bench [19]—with further full workflow tests on eight real-world cases. A common limitation of these works is their emphasis on introducing new methodologies without delving into why such agents perform effectively or critically examining the current constraints of bioinformatics agents. While some efforts aim to establish benchmarks for bioinformatics agents [14,15,16,17,18,19,20], they still suffer from critical limitations (Additional file 1: Fig. S1). First, the task coverage remains narrow and lacks technical depth. For example, GenoTEX [17] is limited to gene data expression analysis while ScienceAgentBench [20] offers cross-disciplinary but shallow scenarios. Their task formats are confined to simplistic formats (e.g., question answering, multiple-choice, or modular sub-tasks) with some even omitting code execution. This lack of sufficient and deep task coverage forces researchers to expend significant effort verifying agent adaptability across various experimental conditions. Second, current benchmarks typically rely on single-dimension metrics including task success rate, question-answering accuracy, or code execution correctness. Therefore, they fail to quantify core agent abilities, such as agent cognition, collaboration, efficiency, knowledge integration or others. Third, existing benchmarks often rely on non-standard, closed and task-specific evaluation structures, lacking compatibility with diverse agent frameworks, large language models (LLMs), and execution languages, which hinders reproducibility and comparative evaluation. For instance, assessments based on QA or multiple-choice cannot be directly applied to a researcher's real-world task without first artificially adapting it to specific formats. Fourth, most benchmarks merely report results without conducting diagnostic analysis—such as investigating why agents fail specific tasks or determining which factors influence their ability to automate bioinformatics tasks.

In this study, we introduce a comprehensive and standardized benchmarking evaluation system for agent-based single-cell omics analysis, facilitating the field's transition from experience-driven approaches to an agent-ecosystem paradigm. This study delivers four core innovations: (1) An open-source agent benchmarking platform with standardized interfaces compatible with agent frameworks (e.g., ReAct [21], LangGraph [22], AutoGen [23]), supporting integration of diverse LLMs (e.g., GPT-4o [24], GPT-4.1 [25], DeepSeek-R1 [26], DeepSeek-V3 [27], Qwen-2.5-max [28], Sonnet-3.7 [29], Gemini-2.5-pro [30], and Grok3-beta [31]) and programming language (e.g., Python and R). This platform serves as a unified evaluation platform adaptable to various single-cell omics analysis tasks and enabling combinations across agent frameworks and LLMs; (2) Multidimensional evaluation metrics comprehensively evaluating core agent capabilities with 18 elaborate metrics across four aspects: Cognitive Program Synthesis, Collaboration and Execution Efficiency, Bioinformatics Knowledge Integration, and Task Completion Quality; (3) Integrated 50 benchmarking tasks of single-cell omics analysis—each employing core analytical tools with public datasets—spanning diverse task types, species, omics, programming languages, and sequencing technologies; (4) Attribution analyses on result consistency/biological metrics, agent robustness, functional modules and task failures to explore key factors that influence agent capabilities in single-cell omics analysis. The proposed benchmarking evaluation system facilitates an objective, quantitative comparison of the performance of heterogeneous agents within a unified, reproducible environment. It not only provides a robust empirical foundation for selecting optimal agent practices but also underscores critical agent limitations—specifically in high-quality code generation, long workflow context handling, and accurate context-aware knowledge retrieving. By filling the gap in comprehensive benchmarking, the study provides a critical empirical foundation and best practices for developing robust AI agents in computational biology.

# II. Results

## A. Evaluation system overview

To systematically evaluate the capabilities of LLM-based agents in single-cell omics analysis, we developed a benchmarking evaluation system comprising three essential components, each contributing to a distinct aspect of the evaluation process (Fig.1):

**Benchmarking platform:** The benchmarking platform is built upon standardized inputs (Agent input), a well-defined unified agent-based system (Agent system) and agent-generated results (Agent output), as illustrated in Fig. 1a. Prompts in agent input is standardized by the prompt template including task descriptions, dataset paths, and analysis requirements while the corresponding raw single-cell omics data (e.g., scRNA-seq and spatial transcriptomics) is stored at the right location. Then, the agent system is responsible for implementing omics analysis tasks under a plan-execute-reflect loop and capable of invoking various tools and LLMs in a dynamic, task-adaptive manner. During task execution, agents autonomously orchestrate tool usage, LLM calls, and reasoning steps, with the entire process monitored and logged for downstream evaluation. The final outputs encompass both computational and visual results. The benchmarking platform supports diverse agent frameworks (e.g., single-agent and multi-agent) and LLM integration, enabling objective, quantitative comparison of heterogeneous agents' performance within a unified, reproducible environment.

For robust and systematic comparisons, the benchmarking platform integrates eight popular LLMs—GPT-4o (2024-05-13), GPT-4.1 (2025-04-15), DeepSeek-R1 (2025-01-20), DeepSeek-V3 (2024-03-24), Qwen-2.5-max (2025-01-29), Sonnet-3.7 (2025-02-24), Gemini-2.5-pro (2025-03-25), Grok3-beta (2025-02-18)—selected based on their leading performance in code-intensive tasks (Additional file 1: Fig. S1b). These LLMs are deployed within three representative agent frameworks: ReAct, LangGraph, and AutoGen, chosen for their distinct and prototypical architectural designs. ReAct adopts a single-agent paradigm, while LangGraph and AutoGen represent multi-agent frameworks with coordination mechanisms. (Additional file 1: Fig. S1c).

**Evaluation Metrics:** A suite of 18 quantitative metrics was developed to assess agent performance across four key dimensions: Cognitive Program Synthesis, Collaboration and Execution Efficiency, Bioinformatics Knowledge Integration, and Task Completion Quality, as shown in the first four panels of Fig. 1b. A weighted total score was further calculated to provide a holistic summary of performance across these dimensions. These metrics capture a wide range of capabilities, including output correctness, tool usage appropriateness, response coherence, and execution efficiency. Besides, additional analyses were conducted to evaluate robustness to prompt variations, dataset variations, multiple runs and to elucidate the influence of specific architectural components on task outcomes (last two panels of Fig. 1b), with the same set of evaluation metrics applied to ensure comparability.

**Benchmarking Tasks:** 50 representative single-cell omics analysis tasks were compiled to perform systematic evaluation, as shown in Fig.1c. Each task incorporates a core analytical tool, a real-world dataset, and an analytical script that provides detailed analysis code along with the corresponding gold-standard output results (more details in Additional file 2: Table S1). These tasks can be categorized into distinct types， including batch correction, cell annotation, dynamic analysis, perturbation, ATAC-seq analysis, multi-omics analysis, spatial deconvolution, gene imputation, spatial domain identification, cell-cell communication, clustering, HVGs selection, and other four independent tasks. Encompassing diverse objectives, multiple species, multi-omics data, various programming languages, and multiple sequencing technologies, these tasks enable detailed performance evaluation of LLM-based agents under real-world scenarios and facilitate analysis of their generalization capabilities and robustness across distinct biological and computational characteristics conditions.

## B. Benchmarking of multiple agent frameworks cross diverse LLMs

Using the benchmarking platform illustrated in Fig. 1a, we performed the main experiment to evaluate agent capabilities for automated omics analysis across multiple dimensions, testing three agent frameworks (i.e., AutoGEN, LangGraph, ReAct) and eight widely used LLMs (i.e., GPT-4o, GPT-4.1, DeepSeek-R1, DeepSeek-V3, Qwen-2.5-max, Sonnet-3.7, Gemini-2.5-pro, and Grok3-beta). The main results are presented in Fig. 2a. Vertically, it displays the outcomes for three agent frameworks combined with eight LLMs. Horizontally, it shows results across 18 multidimensional evaluation metrics, where the first 17 metrics are aggregated into a Total Score (ranging from 0 to 1, where higher values indicate better performance). The total score serves as the 18th metric, presented in the final column of the table. The preprocessing of the benchmarking tasks and more detailed information about the vanilla architectures of multiple agent frameworks can be found in Additional file 1: Fig. S2. The prompts of benchmarking tasks can be found in Additional file 2: Table S3. Notably, regarding the evaluation metrics involving LLM scoring, we averaged the scores from three distinct LLMs to mitigate potential biases. Further details (e.g., changes between single-judge and cross-judge scoring, agreement rate between human experts and LLM assessments) are provided in Evaluations of Section IV.D and Additional file 3: Text S1. Besides, in addition to using LLMs and existing similarity evaluation tools (i.e., AST and ROUGE) to assess the code, we also conducted a lightweight execution-grounded check from data preprocessing outputs and two manually defined checks, as shown in Additional file 1: Fig. S3b-e. Further details are provided in Evaluations of Section IV.D and Additional file 3: Text S2.

From the results shown in Fig. 2a, we can make several observations. First, comprehensive evaluation across all metrics identified GPT-4.1 and Grok3-beta as top performers for AutoGEN, Sonnet-3.7 and Grok3-beta for LangGraph, and Grok3-beta and GPT-4.1 for ReAct, with Grok3-beta consistently achieving optimal task success rates across frameworks. Grok3-beta demonstrates the strongest cross-framework adaptability, consistently ranking among the top 2 performers across agent frameworks. To directly compare core LLM capabilities—isolated from any agent framework—we integrated results across frameworks for LLMs. As shown in Fig. 2b, Grok3-beta demonstrates improved performance over GPT-4.1 and Sonnet-3.7 across selected metrics and diverse task categories, particularly excelling in code scores, retrieval accuracy, and task completion rate. Notably, Grok3-beta falls behind Sonnet-3.7 in plan score and behind GPT-4.1 in RAG-triggering accuracy, reflecting distinct strengths among the LLMs. This pattern of relative strengths is further reflected in the per-task ranking distributions for plan scores and code AST similarity (Fig. 2d). We also computed Total Score for each LLM over the 50 tasks and reported total scores across the 50 tasks in Fig. 2c, along with a statistical comparison between Grok3-beta and Sonnet-3.7, and the distribution of tasks in which each LLM ranked within the top 1 to top 8. Second, the ReAct framework paired with Grok3-beta achieved the highest total score and excelled in task completion quality, demonstrating the efficacy of iterative reasoning-action loops in single agent for structured scientific tasks. However, ReAct's complete failure with DeepSeek-V3—which explains why DeepSeek-V3 results were omitted from the ReAct section in Fig. 2a—reveals ReAct's acute dependency on LLMs natural capabilities and weaker LLMs cannot compensate through architectural design alone. In the ReAct framework,

DeepSeek-V3 was unable to properly trigger tool invocation, leading the agent to falsely assume task termination. Third, ReAct necessitates two to three times more interaction and correction rounds compared to LangGraph and AutoGEN. Nevertheless, it achieves 12–18% higher retrieval accuracy on average. Fig. 2e (left) illustrates variations in average interaction rounds and knowledge retrieval accuracy of ReAct compared to multi-agent frameworks across different task categories. Fig. 2e (right) displays the number of tasks in which each framework ranked among the top 3 on knowledge retrieval accuracy. These results indicate that the single-agent advantage arises from superior knowledge-fusion capabilities: ReAct's streamlined, sequential design reduces decision latency and error propagation, making it particularly effective for time-sensitive and accuracy-critical retrieval tasks. However, this comes at the cost of increased cognitive load on a single agent, which can cause more interaction rounds that is a limitation mitigated in multi-agent systems through role specialization. Finally, the total scores of the eight LLMs across different task types and agent frameworks are presented in Additional file 1: Fig. S4 to illustrate their potential task-specific strengths.

Furthermore, we observed a strong positive correlation between code scores in Cognitive Program Synthesis and Task Completion Quality from Fig. 2, indicating that executable code generation is the primary driver of successful bioinformatics automation. Plan scores showed no significant correlation with Task Completion Quality, as evidenced by AutoGEN+DeepSeek-r1 matching top models (Grok3-beta and GPT-4.1) in plan score despite 66% lower task success rates. To validate this phenomenon, we computed correlations between task completion rate and different evaluation dimensions (e.g., plan metrics, code metrics) in Additional file 1: Fig. S5. Specifically, we conducted a correlation analysis across 24 experimental configurations, each comprising a unique combination of three agent frameworks and eight widely used LLMs.

For each configuration, we computed the spearman correlation coefficients between task completion rate and a set of fine-grained evaluation metrics grouped under three categories: planning performance, code generation quality, and knowledge retrieval effectiveness. As shown in Additional file 1: Fig. S5, the results indicate that, in most cases, task completion rate tends to exhibit a stronger positive correlation with code-related metrics, whereas correlations with planning and retrieval-related metrics appear generally weaker and less consistent across LLMs and frameworks. These findings suggest that, under the current experimental setup, the performance of agents in completing tasks depend more critically on their capabilities in code generation, rather than on planning strategies. Task failures often stem from code errors that agents cannot independently resolve, typically originating in data processing stages. Three specific failure examples (Cell2location, Scanorama and NovoSpaRc) are shown in Additional file 1: Fig. S3f.

## C. Analysis of result consistency and biological metric

In Fig. 2a, we evaluate the agent's task performance using result consistency with the ground-truth script—a quantitative metric detailed in the Section V.B. As shown, the highest result consistency barely reaches around 0.4, indicating that most tasks either failed or achieved only limited alignment with the expected outcomes.

To investigate the causes of task failures and low consistency, we selected several representative cases for detailed analysis in Additional file 1: Fig. S6. The figure illustrates how specific coding errors at key steps led to unsuccessful outcomes or reduced alignment with ground-truth results. As an example, Additional file 1: Fig. S6a presents a complete code snippet from the DestVI task, comparing two combinations of agent frameworks and LLMs: AutoGEN+Grok3-Beta, which showed low consistency with ground truth, and AutoGEN+Sonnet-3.7, which achieved higher result consistency. The analysis reveals that a critical error occurred when assigning the labels_key parameter: Grok3-Beta failed to provide the correct value "broad_cell_types", ultimately resulting in inconsistent feature dimensions after training. Additional file 1: Fig. S6b presents several key code blocks to illustrate typical cases of output shape mismatch and low consistency. We observe that shape discrepancies often arise from variations in data preprocessing—such as differences in parameter settings during cell filtering. Beyond preprocessing inconsistencies, low alignment is frequently attributable to critical parameter misconfigurations during model training setup, like the velocity mode specification in scVelo. Additional file 1: Fig. S6c further showcases two representative task failures. In both cases, errors occurred during the data processing stage due to missing essential steps. For example, in the cell2location task, genes were not renamed to ENSEMBL IDs, which is required for correct matching between single-cell and spatial data. This omission led to subsequent failures in model training. Moreover, Fig. 3a visually compares the ground-truth result and the agent-generated output in four analysis tasks. In PAGA, the agent’s result shows fewer graph connections than the ground truth, which stems from its omission of a key graph denoising step—specifically, the sc.tl.diffmap() operation. More visual results are shown in Additional file 1: Fig. S7a for spatial deconvolution and batch correction. Therefore, most task failures and low consistency outcomes to the agent’s lack of familiarity with the structure of input data. This reveals a key limitation in applying current AI agents to bioinformatics tasks: even with self-reflection capabilities, they often fail to adequately handle diverse data preprocessing strategies.

Another important consideration is that result consistency, while reflecting similarity to the ground-truth script, does not inherently speak to the biological quality of the result. To address this, we evaluated a subset of tasks using task-specific biological metrics (Additional file 1: Fig. S7d), which are computed against actual biological labels rather than the outputs of the ground-truth script.

To determine whether result consistency and task-specific biological metrics reflect similar trends, we first compared both measures across four tasks—DestVI, Scanorama, GimVI, and Leiden—in Fig. 3b. The biological metrics (unfilled bars) and result consistency values (shaded bars) show that agent–LLM combinations with high biological performance generally also achieve high consistency, while those with the lowest biological scores typically exhibit the poorest consistency. We further analyzed the correlation between these two metrics across the same tasks in Fig. 3b, where dotted lines indicate weak correlation for some data points. More correlation analysis for cell annotation (CellAssign), perturbation prediction (scGEN), gene imputation (SpaGE) and HVGs selection (Triku) are shown in Additional file 1: Fig. S7b. Most tasks show a positive correlation between consistency and biological metric scores. In Scanorama, although a weak negative correlation is observed, all aggregated scores across agent–LLM combinations remain above 0.6, indicating consistently adequate performance. Finally, we visualize both metrics as heatmaps for this task subset in Fig. 3c and 3d. Here, white squares indicate missing values where either consistency or the biological metric could not be computed. Notably, local trends in color intensity often align between the two maps (highlighted by green boxes). For instance, in the scGEN task, variations in result consistency across Qwen-2.5-max, Sonnet-3.7, and Gemini-2.5-Pro are mirrored by corresponding changes in biological metric scores.

Therefore, while result consistency does not equate to task-specific biological metrics, it still serves as a meaningful proxy for outcome quality. An advantage of using result consistency is that it enables cross-task comparison under a unified scale, facilitating evaluation across diverse datasets and tasks, and allowing systematic assessment of agent capabilities across different task categories. Moreover, when the ground-truth script represents a high-quality solution, result consistency becomes an even more reliable indicator of biological performance.

## D. Robustness analysis across prompts, datasets, and multiple runs

In this section, we investigate the robustness of agents starting from prompt variations. Three distinct prompt tiers are designed (Additional file 1: Fig. S8, Additional file 2: Table S3): (1) Basic Prompts, which contain fundamental task, dataset, and output result descriptions, explicitly structured into three components: Task description, Data location, and Must-save Results (this is the prompt tier used for the results in Fig. 2); (2) Intermediate Prompts, which extend the Basic tier by augmenting them with a Key Requirements component containing key analytical steps or preprocessing steps (e.g., in perturbation tasks, "Calculate cell cycle labels to be used based on the cell cycle genes file" is added into the prompts as one key preprocessing step); and (3) Advanced Prompts, which build upon the Intermediate tier by incorporating a Core Analysis Steps component that provides straightforward planning procedures. To isolate the effect of prompt design, we evaluated all prompt tiers using Grok3-beta, the highest-performing LLM from main experiments. Performance results under the multidimensional evaluation metrics are presented in Fig. 4a.

As shown in the results, AutoGEN and ReAct exhibit marginal improvements (2%-15%) with intermediate/advanced prompts, while LangGraph shows a negligible decrease (~0.01). To explain this opposite trend, we present metric differences on intermediate/advanced prompting strategies across three agent frameworks in Additional file 1: Fig. S9b. The figure shows that LangGraph experiences significant declines primarily in Code Consistency with Plan and Retrieval Accuracy, indicating these capabilities (coding and knowledge retrieval) are potentially critical factors affecting performance. We further speculate that the observed opposite trend also may arise from variations in framework design flexibility. Among the three agent frameworks, ReAct employed the most flexible design, utilizing a single agent with dynamic reasoning. AutoGEN offered intermediate flexibility, introducing a manager agent to autonomously determine the next step and system termination. LangGraph demonstrated the least flexibility, as we implemented deterministic rules to strictly control termination timing and step execution (Additional file 1: Fig. S2b). Thus, enhancing framework architecture flexibility can counteract the inherent prompt sensitivity of LLMs.

Furthermore, compared to basic prompts, both intermediate and advanced prompts demonstrate improved scores in plan and code. However, task success rates showed no significant improvement with intermediate/advanced prompts and even declined for LangGraph, which contrasts with rising plan quality. To visualize the relationship between task success and the metrics Plan Score and Code Consistency with Plan, Fig. 4b uses bubble size and color intensity to represent their values across tasks, frameworks, and prompt levels. We observe that smaller bubbles correlate strongly with task failure. In contrast, lighter color (indicating lower Plan Score) shows no consistent association with failure, as evidenced by successful tasks like Stereoscope and scGEN in LangGraph (advanced prompts) and Hotspot in AutoGen (advanced prompts). In failed tasks using intermediate/advanced prompts, two limitations of current agents emerge: (1) Extended workflows introduce operational instability where agents commit errors in steps they previously executed correctly under basic prompts. (Additional file 1: Fig. S9a (left)). (2) Inaccuracies in knowledge retrieval, particularly when advanced procedures demand precise implementation. This was exemplified in decoupler (Additional file 1: Fig. S9b (right)) under advanced prompts, where agents retrieved incorrect usage information for the Decoupler package in the model training step.

Next, we examined the robustness of agent performance to dataset diversity. We collected an additional dataset for one subtask/tool under each major task category, resulting in two distinct sets of datasets—each comprising 13 tasks (26 datasets in total). The first set corresponds to the datasets used in the main experiments, while the second set consists of newly introduced datasets. Detailed information for both sets is provided in Additional file 1: Fig. S10a and Additional file 2: Table S2. Using the new dataset, we repeated the same experimental procedure and evaluation as in Fig. 2, with consolidated results shown in Additional file 1: Fig. S10b. Overall, the key findings from Section II.B remain consistent: Grok3-beta continues to deliver the strongest overall performance among the LLMs and frameworks; ReAct achieves higher accuracy in knowledge retrieval compared to AutoGEN and LangGraph. These results indicate that agent performance is robust to variations in the dataset. While Additional file 1: Fig. S10b presented combined results from both dataset sets, we further disaggregated the results in Fig. 4c and 4d to compare performance across the two dataset versions for the same tasks. In Fig. 4c, we compare AutoGEN/ReAct+Grok3-beta across eight evaluation metrics using both dataset sets. Although slight fluctuations are observed in most metrics, planning- and coding-related scores (Plan Overall, Code GPT, Code AST similarity) demonstrate greater stability. Fig. 4d illustrates task-level success rates for the two dataset sets (simply denoted as Data1 and Data2). While success rates indeed differ between datasets, the magnitude of change does not alter model rankings—models that originally performed poorly do not become top performers.

Finally, we investigated whether repeated runs under identical prompts, datasets, and environmental configurations except for seed settings would lead to significant variations in results. We selected four LLMs: Grok3-beta, GPT-4.1, GPT-4o, and DeepSeek-R1, representing top-tier (Grok3-beta, GPT-4.1), mid-range (GPT-4o), and lower-performing (DeepSeek-R1) models based on overall benchmark performance. Each model was executed two additional times under the same experimental setup, resulting in three independent runs per model (including the initial run from Fig. 2a), with the only variation arising from the inherent stochasticity in model generation. We computed the deviation of each metric across the three runs, as summarized in Fig. 4e and Additional file 1: Fig. S10d. GPT-series models demonstrated greater robustness, exhibiting smaller fluctuations across runs in execution time, code score, and other metrics. Although minor variations were observed in individual metrics, key performance indicators including pass rate, success rate, result consistency, and total score remained largely stable. We further analyzed per-task result consistency across three runs for Grok3-beta and GPT-4.1 in Additional file 1: Fig. S10c. Grok3-beta produced fully consistent outputs with computable consistency across all tasks in every run. For GPT-4.1, the *scpnmf* task became computable for consistency in the second run. Notably, most tasks showed no change in consistency, and any observed variability remained within a 10% deviation. No previously low-performing task exhibited unexpectedly high performance in subsequent runs.

In summary, variations in input prompts influence agent performance most, while the agent demonstrates robustness to both dataset diversity and repeated experimental runs.

## E. Impact of functional modules within agent frameworks

To evaluate the contribution of distinct functional modules within agent frameworks, we conducted ablation studies on two prompt-robust frameworks (i.e., ReAct and AutoGEN) and one best LLM (i.e., Grok3-beta). Four functional ablations were implemented (Additional file 1: Fig. S11a): (1) w/o retrieve: disabling tool-based knowledge retrieval in the coder agent, restricting task execution to internal knowledge; (2) w/o planning: eliminating explicit planner agents and planning-related prompts, enabling one task execution agent (the single agent itself in ReAct) to autonomously determine actions, RAG invocation, and process termination; (3) w/o reflection: deactivating the coder agent's self-correction capability, causing immediate system termination upon execution errors; (4) w/o workflow control: removing inter-agent communication constraints and replacing predefined speaking orders with dynamic speaker selection by a group administrator. Given that ReAct inherently lacks removable reflection modules and inter-agent communication mechanisms, ablations (1) and (2) were applied exclusively to ReAct, while AutoGEN underwent full ablations (1) – (4). The system prompts for AutoGEN and ReAct can be found in Additional file 2: Table S7 and S8. Experimental results are illustrated in Fig. 5a. AutoGEN and ReAct in the figure correspond to the non-ablated versions.

Analysis of the evaluation results reveals several key findings. First, removing external tools caused performance declines in both AutoGEN and ReAct, reflecting current LLM limitations in bioinformatics knowledge. As shown in Additional file 1: Fig. S11b, AutoGEN showed smaller reductions (task success rate: -0.14; total score: -0.054) than ReAct (task success rate: -0.26; total score: -0.113), indicating the multi-agent framework reduces reliance on RAG. This suggests collaborative agents can partially compensate for specialized knowledge gaps. Second, disabling the requirement for agents to plan before acting led to decreased performance for AutoGEN, while ReAct showed the opposite trend (Additional file 1: Fig. S11b). This demonstrates AutoGEN's stronger reliance on explicit planning compared to ReAct. Furthermore, eliminating planning significantly increased the agent interaction rounds and self-correction rounds, highlighting that planning can improve agent collaboration efficiency and reduce communication overhead. Third, disabling the agents' self-reflection capability resulted in a significant performance decline on the single-cell omics analysis tasks. In Additional file 1: Fig. S11c, we present the task completion rates for AutoGEN and AutoGEN w/o reflection across benchmarking tasks, ordered from highest to lowest completion rate (left to right). The results show that even for relatively simple tasks like scvi and harmony (used for cell integration), the agents struggle to complete the tasks fully without self-reflection. This underscores the critical role of self-correction in maintaining accuracy and robustness during complex task execution through real-time error detection and correction. Fourth, removing the internal workflow control mechanism (i.e., allowing agents to speak freely without orchestration) did not alter the final task success rate and produced a marginal increase in overall score (~0.011). This implies that with well-designed system prompts, agents can effectively self-organize and complete tasks according to their designated roles without explicit turn-taking directives.

Given the significant increase in average interaction rounds observed without planning in both two-agent frameworks, we analyzed 50 benchmarking tasks using three metrics: average rounds (normalized to [0,1], higher values indicate fewer rounds), code quality (mean of code attribute, AST similarity, and ROUGE scores), and task success status (binary variable: 0 denotes failure, 1 denotes success). Fig. 5b shows the results, with bubble size representing code quality, color intensity indicating average rounds, and line type denoting task completion status. Results demonstrate that ReAct consistently required more collaboration rounds than AutoGEN across all conditions, underscoring the inherent advantage of multi-agent systems in enhancing coordination efficiency. Crucially, disabling pre-task planning triggered dramatic increases in interaction rounds for both frameworks, indicating that planning prevents redundant iterations by establishing clear execution pathways. In w/o reflection setting, the immediate termination of task execution upon encountering any code error led to a high rate of task failures. The Hotspot task exemplifies the consequences of removing functional modules: whereas the original AutoGEN and ReAct workflows succeeded, disabling the retrieval module caused parameter-passing errors during Hotspot initialization (e.g., ValueError: Neither `latent_obsm_key` or `tree` or `distances_obsp_key` arguments were supplied).

Analysis conclusively demonstrates that the agent self-reflection mechanism provides the most substantial improvement to framework performance, because it enables real-time error correction and adaptive strategy refinement. RAG augmentation constitutes the second critical enhancement, attributable to its capacity to integrate external bioinformatics knowledge for context-aware decision-making. The impact of workflow planning is framework-dependent: it improves results for AutoGEN, presumably due to its structured architecture benefiting from predefined action plans, while diminishing performance for ReAct possibly because constrained planning disrupts its dynamic reasoning. Explicit inter-agent workflow control exerts only a minimal impact on overall outcomes, suggesting that clear system prompts enable agents to self-coordinate effectively.

## F. Analysis of failed tasks

To gain deeper insight into the failure mechanisms of agent systems when performing complex tasks, we conducted a systematic error-type analysis on failure tasks observed in the main experiments. By analyzing execution logs across combinations of three agent frameworks and eight language models — excluding DeepSeek-V3 under ReAct due to its complete failure, which was consistently omitted in all downstream analyses—we identified several error types that occurred with notably higher frequency (Additional file 2: Table S9).

In particular, inconsistent planning behavior, planner issues, poor instruction following, ambiguous input prompts, and knowledge acquisition failures emerged as dominant categories. The distribution of these high-frequency error types varied across frameworks, suggesting that system architecture may influence error patterns. For instance, errors related to knowledge acquisition failures and redundant process or code were more prevalent in the LangGraph framework, whereas ReAct was more susceptible to long context handling failures (Fig. 6a). Further analysis at the task level revealed that specific tasks—such as cell2location, scMoMaT, gimVI, CellPhoneDB, and NicheNet—tended to exhibit higher failure rates, indicating that their inherent complexity posed greater challenges to the agents (Additional file 1: Fig. S12).

Given the potential structural impact of error types on output quality, we next assessed the performance differences associated with the top eight most frequent errors across several evaluation metrics. These eight error types were selected for having at least 80 instances each, ensuring statistical reliability. Specifically, we compared the score distributions of task instances with and without each error type. The results indicate that instances without these errors generally achieved higher scores in code metrics, code consistency with plan, and task completion rate. In contrast, no significant differences were observed in planning metrics or knowledge retrieval metrics (Fig. 6b). These findings suggest that the impact of frequent error types may be more pronounced in the downstream stages of code generation and code consistency with plan, rather than in early-stage planning or knowledge retrieval performance.

To more systematically investigate the relationship between error distribution and agent performance, we calculated Spearman correlations between the frequency of error types and evaluation metrics. The results revealed significant negative correlations between the distribution of the error types and code quality metrics, code consistency with plan, and task completion rate (Additional file 1: Fig. S12). These correlations support the hypothesis that task failures may stem from the disruption of code coherence and structural stability, ultimately impairing task execution.

In a further grouped correlation analysis, we explored differences in metrics interdependencies between samples with and without each error type. For coder issues, the positive correlation between task completion rate and code metrics was markedly stronger in the error-absent group, suggesting that the absence of such errors allows code quality to more reliably support task completion. Notably, a stronger correlation with code consistency with plan was observed in the error-present group, indicating that this error type may still affect structural coherence under certain conditions. The most prominent differences emerged for long context handling failures, where samples without the error consistently showed stronger positive correlations among task completion rate, code quality, and code consistency with plan, indicating that failures in long-context handling may lead to broad, cross-dimensional degradation (Fig. 6c).

Given these observations, we conducted a focused analysis on long context handling failure to better understand its potential mechanisms of impact. Among samples exhibiting this error, we observed significant differences from the error-absent group across both joint and marginal distributions of key metrics, including code consistency with plan vs. task completion rate, code consistency with plan vs. code quality, and code quality vs. task completion rate. In general, the error-absent samples were more likely to achieve higher scores across all three metrics, whereas the error-present group exhibited a downward trend in overall performance (Fig. 6d). These findings suggest a plausible underlying mechanism: deficiencies in long-context handling may compromise the agent's ability to maintain alignment with the original plan, resulting in cumulative deviations during code generation. Prior studies [32,33] have shown that LLMs exhibit strong positional biases—favoring content at the beginning and end of context while exhibiting reduced sensitivity to middle-span information, even when it is task-relevant. As a result, middle-span contextual content may be underutilized during generation, disrupting the alignment between planned intentions and executed code. These disruptions extend beyond isolated syntactic or logical flaws; instead, they tend to propagate across the code generation process, compounding into broader structural and semantic inconsistencies. Ultimately, such degradation undermines functional completeness, compromises logical rigor, and diminishes execution efficiency—leading to cascading failures that reduce overall task success.

In summary, high-frequency agent errors are linked to declines in code quality, plan-code consistency, and task completion. Long-context handling failures, in particular, show broad cross-metric impact and may trigger cascading effects by disrupting plan adherence, which in turn impairs code quality and task success. Correlation patterns suggest that plan-code consistency may mediate this degradation process.

# III. Discussion

This work establishes a comprehensive benchmarking evaluation system tailored for agent-based single-cell omics analysis, with the aim of evaluating the effective application of agent systems in the field of bioinformatics. This work presents three primary contributions. First, we provide empirically grounded guidance for selecting appropriate LLM-agent combinations for single-cell omics analysis, revealing significant performance variations across LLMs and agent frameworks. Second, we offer actionable insights for agent *design* and *optimization* through a series of attribution analyses. The ablation studies quantified the functional impact of core agent modules; prompt robustness analysis identified effective engineering strategies; and failure task analysis pinpointed critical failure modes, laying a foundation for building more reliable bioinformatics agents. Third, this benchmarking evaluation system serves as a methodological blueprint for extending agent capabilities into increasingly complex biological computing scenarios, with the potential to significantly automate experimental workflows and accelerate discovery.

Despite establishing this benchmarking evaluation system, several limitations need careful consideration and highlight avenues for evolution. First, the scope of applicability in bioinformatics. Our evaluation, while covering 50 high-frequency workflows, inevitably faces the rapid evolution inherent to single-cell biology. Lower-prevalence yet practically relevant long-tail workflows and emerging technologies (e.g., ultra-high-resolution spatial transcriptomics) are not represented. Besides, our scope is currently limited to three well-established, widely adopted agent frameworks, excluding niche or emerging systems with smaller user bases or higher technical barriers. Maintaining the benchmark's relevance demands a proactive, community-driven approach for continuous integration of new tasks, new LLMs and new agent frameworks. Second, the black-box of agent decision-making presents a significant hurdle to biological trust. While our quantitative metrics assess task success, they do not inherently guarantee the biological plausibility or mechanistic interpretability of the agent's internal reasoning. Building biologist confidence necessitates the future development and integration of explainability tools (e.g., attention visualization, intermediate reasoning trace analysis grounded in biological knowledge graphs) alongside core performance metrics. Third, more comprehensive system robustness analysis. While we incorporated diverse task types and agent frameworks to mitigate variability in language model behavior, our current robustness evaluations are constrained to controlled perturbation scenarios, do not yet address more complex challenges such as out-of-distribution generalization, adversarial robustness, or adaptation to dynamic environments. Fourth, the practical computational cost of deploying sophisticated agents, particularly those requiring multi-round LLM interactions, presents a potential barrier to widespread adoption. In this work, we only report the total expenditure for each LLM and agent framework combination on the benchmarking tasks (Additional file 2: Table S11) while per-task token counts, latency, and cost were not logged. In real-world deployment, however, the time and financial overhead associated with running these agents at scale must be rigorously quantified and mitigated through optimizations (e.g., model distillation, efficient prompting, local LLM alternatives) to ensure accessibility and sustainability.

Building on the established framework and its current limitations, our study reveals several insights suggesting promising future directions. To advance this field, future work must: (1) develop structure-aware diagnostic evaluation frameworks that dynamically adapt tasks, metrics, and prompts for equitable agent comparison; (2) leverage behavioral indicators—such as generalization and compositionality—on goal-conditioned datasets spanning diverse environment dynamics to probe the quality of implicit world models encoded within agent policies [34]; and (3) rigorously validate agents in complex biomedical domains (e.g., drug discovery) to assess cross-domain reasoning. Crucially, translating agents into robust scientific tools requires co-designing hybrid human-AI workflows for domains like single-cell omics analysis. This involves addressing key bottlenecks in code generation, long-context handling, and context-aware knowledge retrieval to achieve reliable integration, where biologists guide objectives and agents handle execution. These structure-aware, behavior-based diagnostics are vital for iteratively refining trustworthy scientific AI assistants.

# IV. Conclusions

We developed a comprehensive benchmarking system for evaluating agent-based single-cell omics analysis, integrating a unified platform that is compatible with multiple LLMs and agent frameworks, multidimensional evaluation metrics, and diverse real-world tasks. This study provides support in three key aspects: (1) empirically grounded guidance for selecting LLM-agent combinations in bioinformatics, (2) actionable insights for agent design and optimization through detailed capability analysis, and (3) a methodological blueprint for extending agent-based automation to increasingly complex biological computing scenarios—ultimately promising to streamline workflows and accelerate scientific discovery.

# V. Methods

## A. Benchmarking platform

**Knowledge Base Construction for Bioinformatics Omics Tools.** To construct the knowledge base (Additional file 1: Fig. S2a. Dataset construction procedure), we systematically collected publicly available resources from the internet, including web content introducing bioinformatics tools and algorithms, as well as relevant scientific literature. The raw data were subjected to a rigorous preprocessing pipeline, primarily involving relevance filtering and quality control, to improve thematic relevance and ensure data reliability and consistency. The curated data were then transformed into a semi-structured format to facilitate downstream processing while preserving information integrity. Each knowledge unit was standardized in Markdown format and semantically segmented using the MarkdownTextSplitter module from the LangChain [35] framework to retain contextual coherence as much as possible. Based on this processed content, we extracted key metadata fields such as method name, method type, and method description, and constructed indices to support efficient information retrieval. Finally, the structured text units were embedded into 3072-dimensional semantic vectors using OpenAI's text-embedding-3-large model and stored in a Chroma [36] vector database, enabling semantic-level similarity search and query functionalities. To prevent the knowledge base from containing any ground-truth codes or reference solutions, we included only the usage instructions of bioinformatics tools—such as function descriptions and parameter specifications—without providing actual executable code.

**AutoGEN.** The multi-agent system built with the AutoGEN framework (Additional file 1: Fig. S2b. Detailed architecture of the agent framework) comprises several agents: a Planner, a Coder, an Executor (equipped with Python and R kernels), a Task Manager, and a retrievable bioinformatics omics tool documentation library. Upon receiving the input, i.e., task, dataset and output result descriptions, the Planner generates an initial plan. The Task Manager then decomposes this plan into sequential steps and manages their iterative execution. Crucially, the Task Manager also determines whether to invoke the Retrieval-Augmented Generation (RAG) function for the current step. If RAG is invoked, the Coder first retrieves relevant information from the documentation library before generating code; otherwise, the Coder generates code directly based on its own knowledge. The generated code is executed by the Executor. If execution succeeds without errors, the Task Manager determines the next step and potential RAG invocation. If errors occur, the Coder modifies the code, which is then re-executed. This iterative loop continues until all steps are completed. Finally, the Task Manager automatically confirms task completion and terminates the system. A maximum of 5 reflection attempts is set for execution at each step, and the overall workflow is capped at 50 execution iterations.

**LangGraph.** The multi-agent workflow built on the LangGraph framework (Additional file 1: Fig. S2b. Detailed architecture of the agent framework) comprises several collaborating agents and tools, including a planner agent, code generation agent, code execution agent, retrieve call agent, and a knowledge retrieve tool for external information access.

The workflow is initiated by the planner agent, which generates a detailed execution plan based on the user-provided input prompt. The system then proceeds step-by-step through the planned tasks, with each step entering an iterative loop involving retrieval decision, code generation, and code execution. Specifically, at each sub-task, the retrieve call agent evaluates whether external knowledge retrieval is necessary based on the current planning context. If so, it invokes the knowledge retrieve tool, which incorporates a basic quality control mechanism to filter out low-quality content and retain only relevant and reliable information.

If retrieval is successful, the retrieved content, along with the current sub-task plan, is passed to the code generation agent to produce executable code. The generated code is then executed by the code execution agent. If execution succeeds, the workflow advances to the next step. In case of failure, the error message is returned to the code generation agent to initiate reflection and correction, repeating until the code runs successfully or a maximum retry limit is reached. To constrain runtime and resource usage, each sub-task allows up to 5 reflection attempts, and the overall workflow is capped at 50 execution iterations.

A task is considered successfully completed only if all planned steps are executed without unresolved errors. If any step fails and cannot be recovered within the specified limits, the task is marked as failed.

**ReAct** (Additional file 1: Fig. S2b. Detailed architecture of the agent framework) adopts a single-agent architecture that directly manages and invokes external tools, including a knowledge retrieval tool and a code execution tool. The agent is responsible for both tool invocation and result handling, operating under the core mechanism of a "thought – action – observation" loop. Upon receiving an input prompt, the ReAct agent enters this iterative process, which continues until all necessary operations are completed and a final output is generated, or the agent determines that the task objectives have been met and terminates the process.

In terms of workflow control, the ReAct agent first generates a detailed task execution plan based on the input prompt. It then proceeds step-by-step through the plan, with each subtask triggering an internal loop consisting of optional knowledge retrieval, code generation, and code execution. If an error occurs during code execution, the agent reflects on the error message, regenerates the code, and re-executes it until the current step succeeds before moving to the next.

While the overall execution logic is similar to that of multi-agent systems built on the LangGraph framework, ReAct consolidates the responsibilities of multiple agents into a single entity. Due to the flexibility inherent in the ReAct architecture, it is not feasible to set an explicit upper limit on the number of reflection-revision attempts; however, to manage computational cost, we impose a maximum of 50 iterations for the entire task loop.

Task completion is not determined solely by the success or failure of individual code execution steps. Instead, the ReAct agent makes autonomous judgments about whether the overall objectives stated in the input prompt have been met. If so, the agent terminates the process successfully; otherwise, if the task goals are not achieved within the iteration limit, the task is considered failed. Compared to LangGraph, which relies on explicit flow control and state transitions to evaluate task success, ReAct offers greater flexibility at the cost of reduced structural control.

## B. Evaluation metrics

To comprehensively evaluate the performance of various agent frameworks and LLMs in single-cell omics analysis, we propose multidimensional evaluation metrics comprising 18 elaborate metrics. These evaluation metrics systematically assess agents' capabilities in independently performing single-cell omics analysis tasks across four key aspects: *Cognitive Program Synthesis*, *Collaboration and Execution Efficiency*, *Bioinformatics Knowledge Integration*, and *Task Completion Quality*. The following paragraphs detail the proposed multidimensional evaluation metrics.

**Cognitive program synthesis.** It assesses agents' ability to decompose bioinformatics analysis problems into logical, executable analysis workflows. This aspect evaluates the core reasoning process where the agent formulates an appropriate step-by-step plan and translates it into valid, functional computational code. Therefore, the plan score measures the logical coherence, biological relevance, and completeness of the proposed analytical strategy, while the code score evaluates the correctness, executability, and adherence to bioinformatics best practices of the generated programming scripts.

Specifically, for the plan score, we define ***Plan Content* score** as a metric that assesses whether a plan generated by agents addresses the four essential stages, i.e., data preprocessing, model building and training, model evaluation and post-analysis; ***Plan Attribute* score** evaluates agent-generated plans across five attributes including clarity, comprehensiveness, structural quality, level of detail, and technical feasibility. To holistically evaluate the plan, we define ***Plan Overall* score** as the average of the plan content and attribute scores. To mitigate bias in LLMs, three LLMs (i.e., GPT-4o, Gemini-2.5-pro, Grok3-beta) are adopted for these plan scores with carefully designed prompts. The average of these scores is then reported in the result figure. Denote *Plan*

*Content* score, *Plan Attribute* score, *Plan Overall* score, plan generated by agents, plan content scoring prompt, and plan attribute scoring prompt as $s_{p_{content}}, s_{p_{attibute}}, s_p, a_p, P_{p_{content}}, P_{p_{attibute}}$, then:

$$s_{p_{content}} = ScoreLLM\left(a_p, P_{p_{content}}\right) \tag{1}$$

$$s_{p_{attribute}} = ScoreLLM\left(a_p, P_{p_{attribute}}\right) \tag{2}$$

$$s_p = \frac{s_{p_{content}} + s_{p_{attribute}}}{2} \tag{3}$$

For the code assessment, we define ***Code Attribute*** **score** based on five properties including key segment matching degree, efficiency, completeness, readability, and logicality. We evaluate agent-generated code against these attributes. Similarly, three LLMs (i.e., GPT-4o, Gemini-2.5-pro, Grok3-beta) are adopted for scoring code attributes through the specific prompt. The average of these scores is then reported in the result figure. Denote code scores, code generated by agents, and scoring prompt as $s_c, a_c, P_c$, then:

$$s_c = ScoreLLM(a_c, P_c) \tag{4}$$

While the agent-generated code is evaluated by GPT-4o across the five attributes, we further quantify the structural equivalence between the Abstract Syntax Trees (AST) of agent-generated codes and ground-truth codes and measuring how closely their underlying logic and implementation align using ***Code AST Similarity*** [37]. We employed the standard ast package to directly obtain the code structure trees for both the agent-generated code and the ground-truth code. The similarity between these trees was then calculated using the *difflib* tool[1]. Denote *Code AST Similarity* as $s_{ast-sim}$, then it can be computed by

[1] https://docs.python.org/3/library/difflib.html

$$s_{ast-sim} = AST(r_c, a_c) \quad (5)$$

Furthermore, we employ the ***Code ROUGE-L* score** [38,39], to measure content overlap between generated codes and reference codes by calculating the longest common subsequence (LCS). Similarly, we used the standard rouge package to directly compute the ROUGE scores of the agent-generated code against the ground-truth code. Denote *Code ROUGE-L* score as $s_{rouge-L}$, then it can be computed by

$$s_{rouge-L} = Rouge(r_c, a_c) \quad (6)$$

These defined metrics provide a multifaceted evaluation of agents' cognitive synthesis capability. They assess not only the correctness and executability of the final output (code) but also critically evaluate the quality and soundness of the underlying reasoning process (plan), which offers significant interpretability into agents' problem-solving behavior. Consequently, plan and code metrics enable a deeper understanding of why and how agents arrive at a specific bioinformatics solution, moving beyond mere output validation to assess the core cognitive processes involved.

**Collaboration and execution efficiency.** It measures agents' effectiveness in performing single-cell omics analysis tasks through collaboration and resource consumption. Key metrics include execution time, resource consumption, total interaction rounds, and self-correction frequency, collectively reflecting the agent's ability to minimize operational overhead and user dependency. Specifically, we define ***Execution time*** denoted as $t$, measuring the overall execution time of the agent system; ***CPU/GPU utilization (%)*** denoted as $s_{ru}$, measuring the usage of CPU and GPU in the agent system; ***Average collaboration rounds***, reporting the average total rounds. Denote average interaction rounds, total interaction rounds and total task steps as $a_{avgtr}$, $a_{tr}$ and $N$:

$$a_{avgtr} = \frac{a_{tr}}{N} \tag{7}$$

***Average self-correction rounds***, reflecting agents' ability to correct errors while performing tasks. Denote average self-correction rounds and self-correction rounds as $a_{avgscr}$ and $a_{scr}$:

$$a_{avgscr} = \frac{a_{scr}}{N} \tag{8}$$

Finally, to evaluate whether the coder adheres to the plan during code writing, we define the ***Code Consistency with Plan***, quantifying the implementation fidelity of the agent-generated code to the agent-generated plan. Three LLMs (i.e., GPT-4o, Gemini-2.5-pro, Grok3-beta) are adopted for scoring the metric through specific prompts. The average of these scores is then reported in the result figure. Denote *Code Consistency with Plan* and the scoring prompt as $s_{ccp}$ and $P_{ccp}$, respectively, then:

$$s_{ccp} = ScoreLLM\left(a_c, a_p, P_{ccp}\right) \tag{9}$$

In summary, these metrics collectively assess agent proficiency in collaborative bioinformatics analysis. Execution time and CPU/GPU utilization directly measure operational efficiency and resource optimization. The average collaboration rounds quantify collaborative effectiveness and the ability to minimize unnecessary coordination overhead. The average self-correction rounds indicate the autonomous problem-solving capability and capacity to independently identify and resolve errors. The code consistency with plan indicates the agent's execution fidelity and capacity to rigorously translate plans into executable solutions.

**Bioinformatics knowledge integration.** It measures agents' ability on unifying bioinformatics analytical tools, and domain expertise to enable intelligent interpretation of analysis tasks. To rigorously assess such ability, we define two key performance indicators including: ***RAG*-triggering *Accuracy***, measuring precision in triggering RAG operations at appropriate times. We employ three LLMs (i.e., GPT-4o, Gemini-2.5-pro, Grok3-beta) for assessing the correctness of the agent's RAG-triggering decisions using specific prompts. The average of these scores is then reported in the result figure. Besides, we selected 13 task subsets and incorporated expert human evaluation for three scoring LLMs. Finally, RAG-triggering Accuracy was computed as a weighted combination: 0.8 × LLM score + 0.2 × expert score, thereby further reducing model-induced bias. Denote rag-triggering accuracy, the scoring prompt, and RAG steps as $s_{rag-t}, P_{rag-t}, l$:

$$s_{rag-t} = 0.8 * \frac{JudgeLLM(l, P_{rag-t})}{l} + 0.2 * expert\ score \tag{10}$$

where $JudgeLLM(l, P_{rag-t})$ obtained the number of steps that should trigger RAG.

***Retrieval Accuracy***, evaluating the relevance of context retrieved by RAG for the task execution. We employ a document matching algorithm [40] to assess the accuracy of the relevance between the retrieved text segments and the right tool documentation in the current task. Denote retrieval text by agents and the reference tool text as $a_{re}, r_t$:

$$s_{ra} = 1 - \frac{EditDistance(a_{re}, r_t)}{\max(len(a_{re}), len(r_t), 1)} \tag{11}$$

where $EditDistance(\cdot)$ represents minimum character edits (add/delete/change) needed to make $a_{re}$ match $r_t$. These metrics directly quantify the robustness of integrated knowledge deployment in automated bioinformatics pipelines.

**Task completion quality.** Beyond assessing agent performance during execution, we critically evaluate the task completion quality. To this end, we define four key metrics: ***Task Completion Rate***, measuring the proportion of planned steps successfully executed as the following:

$$s_{TCR} = \frac{N_{completed}}{N} \tag{12}$$

where $s_{TCR}$ represents Task Completion Rate, $N_{completed}$ represents the completed plan steps number and $N$ represents the total plan step number. Task Completion Rate indicates the agent's ability to progress through the full analytical workflow. ***Passing Rate*** [41] denoted as $s_{PR1}$, was calculated as the number of benchmark tasks passed on the first run divided by the total number of benchmark tasks (i.e., 50 single-cell omics analysis tasks). This metric primarily reflects the agent's reliability and robustness in task execution under deterministic conditions. ***Success Rate*** denoted as $s_{SR}$, a stricter measure than Passing Rate, reporting the percentage of tasks where the agent generated the intended, correct result output. It can be calculated as the number of benchmark tasks that generate the expected result output divided by the total number of benchmark tasks (i.e., 50 single-cell omics analysis tasks). Success Rate specifically evaluates output correctness and objective alignment, whereas Passing Rate assesses first-attempt executability. ***Result Consistency*** denoted as $s_{RC}$, quantifying the agreement between the results generated by the agent and the ground-truth result. Result Consistency is crucial for verifying the accuracy and reliability of the agent's final outputs. We employ three strategies to compute result consistency between agent outputs and ground-truth results based on result types. For gene/annotation/ligand–receptor list outputs, such as selected highly variable genes (HVGs), we compute agreement using a list-matching algorithm. For vector- or matrix-formatted representations (e.g., embeddings), we directly calculate cosine similarity [42] between the agent's result and the ground truth. When representations have incompatible dimensions, we first project both outputs into a common low-dimensional space using Principal Component Analysis (PCA) [43], and then compute the Jensen-Shannon Divergence (JSD) between the two vector representations of the same dimension after PCA to assess consistency. Denote the results generated by agents and the ground-truth results as $a_{result}$ and $r_{result}$, then the formulation is as follows:

$$s_{RC} = \begin{cases} Match(a_{result}, r_{result}) \\ CosineSimilarity(a_{result}, r_{result}) \\ JSD\big(PCA(a_{result}), PCA(r_{result})\big) \end{cases} \tag{13}$$

**Total Score**. Finally, we define a ***Total Score*** derived from the aggregation of the preceding 17 evaluation metrics, yielding a normalized value between 0 and 1. All non- [0,1] scores are first mapped to [0, 1]. Scores generated by LLM are divided by the maximum score value 5. Execution time uses modified logarithmic as follows:

$$s_t = 1 - \frac{log_b(t + c)}{log_b(T + c)} \tag{14}$$

where base *b*=1.8 (smaller b values flatten the curve), *T*=10,800 seconds (maximum runtime), and *c*=100 (controlling curve shape). CPU/GPU utilization employs exponential decay to enhance sensitivity in low-value regions:

$$s_{ru} = \frac{e^{-i \cdot s_{ru}^j} - e^{-i \cdot max^j}}{1 - e^{-i \cdot max^j}} \tag{15}$$

where we set $i$ as -0.03, $j$ as 1.5, $max$ as 10. For both average interaction rounds and average correction rounds, we adopt an exponential decay projection with power scaling:

$$s_{tr/scr} = \frac{e^{-k \cdot s_{tr/scr}^{m}} - e^{-k \cdot max^{m}}}{1 - e^{-k \cdot max^{m}}} \tag{16}$$

where $s_{tr/scr}$, denotes the actual average number of rounds (interaction or correction), max is the maximum average number of rounds (50 for interaction, 10 for correction), $k$ is the decay coefficient controlling the overall decay rate, and m is the power parameter controlling discrimination at low values. We set $k = 0.005$ and $m = 1.5$. For the parameter settings of the mapping function applied to metrics such as time, CPU/GPU utilization, and collaboration rounds, we configure the function to exhibit steeper changes within intervals of concentrated data distribution and gentler changes within sparse regions (e.g., near the maximum value). This configuration achieves greater differentiation in the mapped scores. Then, the final score is computed as follows:

$$S = \omega_1 \frac{s_p + s_c}{2} + \omega_2 \frac{s_t + s_{ru} + s_{tr} + s_{scr}}{4} + \omega_3 \frac{s_{rea} + s_{ra}}{2} + \omega_4 \frac{s_{TCR} + s_{PR1} + s_{SR} + s_{RC}}{4} \tag{17}$$

Where $\omega_1, \omega_2, \omega_3, \omega_4$ represent evaluation weights of cognitive program synthesis, collaboration and execution efficiency, bioinformatics knowledge integration and task completion quality. In this study, we set $\omega_1 = 0.15, \omega_2 = 0.15, \omega_3 = 0.2, \omega_4 = 0.5$. All adjustable parameters are shown in Additional file 1: Fig. S13. We base the weighting of these four components on existing benchmarks and our own empirical observations. Since existing benchmarks[16,18,19] primarily focus on quantitative metrics like task success rate and code success rate, and we also believe that task completion quality (pass rate, success rate, consistency, etc.) best reflects an agent's capability, we assign it the highest weight. Furthermore, since a lack of bioinformatics knowledge and weak coding ability are frequently cited shortcomings in previous work[15,20], knowledge integration receive the next highest weight. Since most metrics in Cognitive Program Synthesis are scored by LLMs, we will assign lower weight to this component, grouping it together with collaboration and execution efficiency.

## C. Benchmarking tasks

In the following, we explain the fundamental principles underlying the 50 single-cell omics analysis tasks used for benchmarking. These tasks were grouped into the following categories: Batch correction, Cell annotation, Dynamic analysis, Perturbation, ATAC-seq analysis, Multomics analysis, Spatial deconvolution, Gene imputation, Spatial domain identification, Cell-cell communication, Clustering, and HVGs selection, along with four independent tasks. Each task employed a publicly available analytical tool configured with public datasets. Comprehensive descriptions — including tool information, tutorial code with parameter settings, and dataset details — are provided in Additional file 2: Table S1.

**Batch correction tasks** involve six state-of-the-art tools: scVI [44] leverages deep generative modeling with variational autoencoders to probabilistically correct batch effects while preserving biological variance, enabling scalable integration of large-scale single-cell datasets. Building on scVI, scANVI [45] incorporates semi-supervised learning to jointly perform batch correction and cell-type annotation using partially labeled data, enhancing biological interpretability during integration. Scanorama [46] employs mutual nearest neighbors and non-linear dimensionality reduction to align datasets in a "panoramic" low-dimensional space, effectively correcting batch effects across heterogeneous cell populations. Harmony [47] iteratively clusters cell and adjusts embeddings via soft clustering and centroid-based correction, enabling rapid and robust integration of diverse datasets without altering gene expression values directly. scPoli [48] utilizes contrastive learning and conditional variational autoencoders to integrate multi-batch data while incorporating biological covariates (e.g., cell type), promoting structure-aware batch correction. scGen [49] applies transfer learning and latent space manipulation via variational autoencoders to generalize across batches and experimental conditions, particularly suited for perturbation-response integration.

**Cell annotation tasks** involve three distinct tools：CellAssign [50] employs a probabilistic graphical model to automatically assign cell types using predefined marker gene sets, enabling robust annotation while accounting for batch-specific noise and unknown cell states. Celltypist [51] leverages logistic regression classifiers trained on extensive reference datasets to rapidly annotate cell types at scale, supporting both pre-trained models and custom references with minimal user input. Decoupler [52] infers cell-type identities indirectly by statistically integrating pathway activity and transcription factor regulons, bypassing gene-level variability to reveal functional annotations from bulk or single-cell data.

**Dynamic analysis tasks** involve four tools: PAGA [53] infers coarse-grained topological trajectories by constructing abstracted graph representations of single-cell data, enabling interpretable modeling of developmental and transition processes without assuming linear paths. scVelo [54] resolves RNA velocity kinetics using a likelihood-based framework that models unspliced/spliced mRNA dynamics, predicting cell-state transitions and latent time without reliance on experimental time series. CellRank [55] combines RNA velocity with transcriptomic similarity via machine learning to compute probabilistic fate maps, identifying transition states, terminal fates, and driver genes along dynamic trajectories. SCORPIUS [56] reconstructs linear or branched trajectories through distance-based algorithms (e.g., minimum spanning trees), prioritizing scalability and robustness for inferring temporal ordering in large datasets.

**Perturbation tasks** involve two deep learning tools: scGEN [49] employs variational autoencoders to model perturbation responses in latent space, enabling prediction of unseen cellular states by transferring learned patterns from observed conditions to novel genetic or chemical perturbations. ContrastiveVI [57] disentangles perturbation effects from confounders using contrastive deep generative modeling, isolating condition-specific responses while explicitly accounting for batch effects and biological covariates.

**ATAC-seq analysis tasks** involve two specialized deep learning tools: PeakVI [58] models chromatin accessibility using a deep generative framework tailored for peak-level data, integrating datasets by denoising counts and correcting technical biases while preserving biological heterogeneity. PoissonVI [59] extends variational inference to bin-level ATAC-seq data with a Poisson likelihood, enabling scalable integration and dimensionality reduction for large-scale chromatin landscapes.

**Multomics analysis tasks** involve five tools: totalVI [60] integrates paired transcriptome and proteome data (e.g., CITE-seq) using a deep generative model that jointly denoises both modalities while correcting technical artifacts in protein measurements. MultiVI [61] unifies scRNA-seq and scATAC-seq data—even with missing modalities—through a multimodal VAE framework, constructing joint embeddings that preserve cross-omics biological relationships. MIRA [62] employs a joint topic modeling architecture on multi-omics inputs to simultaneously infer cell states and regulatory programs, explicitly modeling hierarchical biological structure across modalities. MOFA+ [63] applies factor analysis to decompose multi-omics variance into interpretable latent factors, enabling unsupervised integration of >2 modalities while quantifying feature-specific contributions. Seurat's [64] Weighted Nearest Neighbors (WNN) algorithm integrates multimodal data by constructing joint graphs that balance modality-specific contributions, enabling cluster-free cell state resolution.

**Spatial deconvolution tasks** involve six tools: DestVI [65] extends scVI's generative framework to spatially resolve cell-type proportions and cell-type-specific gene expression within spots by integrating single-cell references and spatial transcriptomics, enabling fine-grained tissue mapping. Stereoscope [66] decomposes spatial spots using negative binomial regression on single-cell references, quantifying cell-type abundances without requiring paired data while maintaining computational efficiency for large datasets. Cell2location [67] employs Bayesian hierarchical modeling to estimate absolute cell-type densities in spatial locations, explicitly accounting for technical variation and providing uncertainty-aware deconvolution of complex tissues. Tangram [68] aligns single-cell profiles to spatial data via deep learning-based optimal transport, reconstructing high-resolution spatial maps of cell distributions and gene expression at sub-spot resolution. GraphST [69] integrates spatial coordinates and transcriptomics through graph neural networks with self-supervised contrastive learning, enabling spatially aware deconvolution that preserves tissue architecture. Novosparc [70] reconstructs spatial tissue organization from scRNA-seq alone using optimal transport principles, predicting spatial gene expression patterns and cell locations without prior spatial data.

**Gene imputation tasks** involve two tools: gimVI [71] leverages a joint generative model to transfer gene expression from scRNA-seq to spatial transcriptomics data, imputing unmeasured spatial genes while integrating shared biological variation across modalities. SpaGE [72] uses transfer learning and optimized k-nearest neighbors to impute high-dimensional spatial gene expression from scRNA-seq references, preserving spatial context with minimal computational overhead.

**Spatial domain tasks** involve five tools: SEDR [73] integrates transcriptomics and spatial coordinates through deep graph autoencoders with variational inference, identifying spatially coherent domains while preserving global tissue architecture and biological context. SpaGCN [74] combines gene expression, spatial location, and histology images via graph convolutional networks to delineate tissue domains with biologically meaningful boundaries and multi-modal interpretability. Squidpy [75] provides scalable spatial neighborhood analysis through graph-based statistics and clustering algorithms, enabling domain segmentation via spatial autocorrelation patterns and community detection in tissue graphs. Hotspot [76] identifies spatially restricted functional domains by detecting local co-expression modules through spatial autocorrelation analysis, revealing niche-specific gene programs without predefined spatial coordinates. Seurat [64] defines spatial domains by integrating transcriptomic similarity and physical proximity through spatially constrained graph clustering, enabling joint analysis with single-cell references for domain annotation.

**Cell-cell communication tasks** involve three tools: CellphoneDB [77] identifies biologically relevant ligand-receptor interactions by integrating curated molecular complexes and statistical permutation testing, quantifying communication probabilities across cell types in spatially unresolved data. NicheNet [78] infers upstream signaling drivers and downstream gene responses using prior knowledge networks, linking ligand activity to transcriptional changes to reveal causal signaling niches. CellChat [79] models communication probabilities via mass action principles and pattern recognition, identifying dominant signaling pathways and global communication architectures across hierarchical scales.

**Clustering tasks** involve two tools: Leiden [80] algorithm optimizes modularity through iterative network refinement and stochastic local moves, enabling high-resolution, scalable community detection in large single-cell graphs while resolving disconnected communities. DouCLing [81] integrates multi-view learning and doublet detection with graph-based clustering, explicitly resolving hybrid transcriptomes to prevent doublet-induced artifacts in cluster assignments.

**HVGs selection tasks** involve six diverse tools: deeptree [82] leverages hierarchical deep neural networks to identify HVGs by modeling gene expression variance across multiple biological scales, prioritizing genes with context-dependent variability relevant to cellular hierarchies. scMoMat [83] integrates multi-omics data through matrix factorization to select HVGs whose variability aligns with cross-modal biological patterns, ensuring consistent feature selection in multimodal studies. Triku [84] employs k-nearest neighbor graphs and persistent homology to distinguish technical noise from biological variability, selecting HVGs that drive meaningful topological structures in single-cell data. SingleCellHaystack [85] detects HVGs using Kullback-Leibler divergence in low-dimensional space, identifying genes with non-random expression distributions across cell states without distributional assumptions. scPNMF [86] applies Poisson-regularized non-negative matrix factorization to model count-based overdispersion, selecting HVGs whose variability exceeds technical expectations in sparse single-cell matrices. Scry [87] implements deviance-based feature selection with generalized linear models, prioritizing genes showing significant excess variability beyond Poisson or negative binomial noise models.

**Other four independent tasks** involve four tools: scdemon [88] identifies cell-type-specific regulatory modules by integrating scRNA-seq and motif accessibility through constrained matrix factorization, revealing context-dependent gene regulation programs across cell states. DESeq2 [89] performs robust differential expression analysis using negative binomial generalized linear models with empirical Bayes shrinkage, enabling precise fold-change estimation and statistical testing for bulk or pseudobulk RNA-seq data. ResolVI [90] enhances spatial transcriptomics resolution through deep generative modeling of spot decomposition, reconstructing subcellular expression patterns by integrating single-cell references and spatial covariance structures. Sctransform [91] normalizes single-cell count data via regularized negative binomial regression, simultaneously removing technical noise and stabilizing biological variance for improved downstream analysis without over-correction.

## D. Experimental implementation

**Main experiment.** In this study, we designed and implemented a unified experimental workflow to evaluate three different agent frameworks in combination with eight mainstream large language models. Specifically, we sequentially submitted a set of 50 single-cell omics analysis tasks prompts (Additional file 2: Table S3. Prompts for 50 single-cell omics analysis tasks) as input to each agent. Upon receiving the input prompt, each agent autonomously initiated its task execution process and recorded comprehensive logs throughout the runtime.

After all tasks were completed, the system automatically triggered an evaluation module to perform quantitative analysis of the execution process. This module incorporated 18 distinct evaluation metrics, each designed to assess performance based on execution logs and final outputs. For each task, key information was automatically extracted from the agent logs using scripts that combine regex-based parsing with GPT-4o assistance to structure the data. Corresponding algorithms, covering all evaluation metrics calculations, were then applied to generate independent evaluation results for each metric. These were subsequently aggregated to produce framework-level model evaluations, enabling systematic comparisons across both agent frameworks and language models. All single-cell omics analysis tasks prompts followed a standardized basic template, which comprised three components: a brief description of the task, the dataset location, and the result-saving path (Additional file 1: Fig. S8).

During task execution, to accommodate the diverse runtime dependencies of different tasks, the system was equipped with an environment-adaptive mechanism. Prior to execution, agents were able to automatically switch to the required virtual environment. For multilingual scenarios, language context was also provided to guide appropriate code generation. In addition, the agents were capable of autonomously identifying input data paths, invoking external knowledge retrieval tools as needed, and saving output files to the designated locations. To ensure computational efficiency and system stability, tasks exceeding predefined runtime thresholds were automatically terminated.

**Robustness experiment.** To further investigate the robustness of agent behavior under varying input conditions, we conducted additional experiments based on three agent frameworks using the Grok3-beta language model as the backend. Specifically, we examined agent performance when handling input prompts of different structural complexity.

Beyond the *basic prompt* used in the main experiments, we designed two additional types of input prompts: the *intermediate prompt* and the *advanced prompt* (Additional file 1: Fig. S8, Additional file 2: Table S3). The *intermediate prompt* extends the basic version by incorporating *key requirements*, which are extracted from detailed analyses of each task’s workflow and emphasize critical components essential for successful completion. Building on this, the *advanced* *prompt* further includes two additional elements: (1) *core analysis steps*, which specify the central data processing or modeling procedures for each task, and (2) *libraries*, which explicitly list the software packages required for execution.

Aside from the differences in prompt structure, the robustness experiments followed the same implementation pipeline as the main experiments. This includes identical task scheduling strategies, logging mechanisms, evaluation metrics, and assessment workflows, ensuring a controlled environment for systematically evaluating the effect of input prompt complexity on agent performance across frameworks.

**Ablation experiment.** To further investigate the contributions of specific functional modules to overall task performance in both single-agent and multi-agent architectures, we conducted a series of ablation experiments using ReAct and AutoGen as representative frameworks. ReAct exemplifies a typical single-agent design, while AutoGen represents a collaborative multi-agent architecture. Given their structural differences, we devised distinct ablation settings tailored to the characteristics of each framework.

For the ReAct framework, which integrates and manages all external tools within a single agent node, we designed two ablation configurations focusing on the knowledge retrieval and task planning components. In the w/o retrieve setting, the knowledge retrieval module—originally integrated into the agent as an external tool—was removed by eliminating the corresponding tool interface. To ensure coherence in the agent's reasoning process, the system prompt was also revised to exclude all instructions and procedural elements related to knowledge retrieval. In the w/o planning configuration, since ReAct does not include a dedicated planning agent as found in the multi-agent framework, planning is implemented through the system prompt of the single agent. To simulate the absence of planning capabilities, we removed all references to planning from the prompt, including step decomposition, action sequencing, and related directives. This setup allowed us to isolate and assess the impact of task planning on the agent's performance.

For the AutoGEN framework, a multi-agent framework, we designed four ablation configurations focusing on the knowledge retreval, task planning, agent reflection and agent workflow control components. In the w/o retrieve setting, similar to the ReAct configuration, the tool-based knowledge retrieval functionality was removed from the coder agent, restricting the agent to rely solely on its internal knowledge for task execution. In the w/o planning setting, we eliminate explicit planner agents and remove all planning-related prompts. A specific agent for task execution (the single agent itself in ReAct) determines the next action, decides when to invoke RAG, and terminates the process. In the w/o reflection setting, we disable the coder agent's self-correction capability. The agent system terminates immediately upon encountering execution errors. In the w/o workflow control setting, we remove restrictions on inter-agent communication. Instead of predefining agent speaking orders, a group administrator dynamically selects the next speaker based on real-time needs.

Illustrations of these ablation configurations are provided in Additional file 1: Fig. S11a. The system prompts for ablation experiments on AutoGEN and ReAct can be found in Additional file 2: Table S7 and S8, respectively.

**Analysis of Failed Tasks.** To systematically identify and categorize critical failure modes occurring during task execution in multi-agent systems, we conducted a comprehensive analysis of all task execution logs from the main experiments. This analysis encompassed combinations of three agent frameworks and eight prominent large language models, with particular focus on tasks exhibiting success rates below 100% under these configurations. Drawing from the consolidation and comparison of these logs, as well as informed by recent systematic studies on multi-agent system failures [92], we distilled fourteen common error types. These were further organized into four high-level categories based on their manifestation and system-level occurrence: system design issues, agent scheduling and collaboration issues, environment and input issues, and core module or model capability issues (Additional file 2: Table S9). This taxonomy forms the foundational framework for analyzing the causes of task failures in our study.

The identified error types span a range of concerns, from architectural-level execution anomalies, through failures in inter-agent collaboration strategies, to challenges related to environmental configuration and input prompt adaptation, as well as potential deficiencies in the core modules' abilities in planning, generation, and execution. To enhance the rigor and reproducibility of error identification and attribution, we developed a dual-stage judgment mechanism combining large-model chain-of-thought reasoning with expert human verification.

In the initial judgment phase, three large language models including Claude 3.7 Sonnet, Grok3-beta and GPT-4.1 were independently assigned to analyse execution logs for each task based on their demonstrated strength and stability in logical reasoning. For each potential error type, the models engaged in structured chain-of-thought reasoning grounded strictly in the original log content. This reasoning required a transparent inference chain detailing possible key triggering factors for the failure, the logical mapping to the candidate error type, and whether critical supporting evidence was present in the logs. This reasoning process served not only as the basis for model-generated judgments but also as a key reference for subsequent expert review, ensuring traceability and logical consistency.

To consolidate model outputs, a voting mechanism was applied whereby an error type was provisionally deemed present if at least two of the three models agreed. Ten percent of the samples underlying these preliminary judgments and their corresponding reasoning chains were randomly selected for review by domain experts to confirm the soundness of the logical inferences and the appropriateness of error attribution.

Furthermore, to mitigate the influence of incidental factors on the analysis, tasks identified as experiencing “interrupted execution” errors were re-executed prior to error classification. Such interruptions typically stem from external system factors—such as network instability, API rate limits, or program crashes—and do not reflect the intrinsic capabilities of the agent systems. Re-running these tasks helped to exclude non-systemic errors induced by environmental variability, thereby improving the robustness of the subsequent failure analysis.

**Evaluations.** Following the calculation method described in Section IV.B, we sequentially computed the results for 17 evaluation metrics, with the total score included as the 18th metric. To mitigate potential biases in evaluation metrics (such as Plan Content and Plan Attribute) when using LLMs for scoring, we employed three distinct LLMs—GPT-4o, Grok3-beta, and Gemini-2.5-pro—and used the average of their scores as the final result. Besides, we observed that LLM-based scoring exhibits larger bias in evaluating RAG-triggering accuracy, with a tendency for models to assign higher scores to their own outputs. To address this, we selected 13 task subsets and incorporated expert human evaluation for three scoring LLMs. The final score of the scoring LLMs was computed as a weighted combination: 0.8 × LLM score + 0.2 × expert score, thereby reducing model-induced bias. The agreement rate between human experts and cross-judge LLM scoring on the selected subset of tasks was approximately 94%. Furthermore, on the full task set, the inter-judge agreement rates among LLMs for the LLM-graded metrics—Plan Content, Plan Attributes, Code Attributes, Code Consistency with Plan, and RAG-triggering accuracy—were 93%, 92%, 95%, 98%, and 90%, respectively. More details are reported in the Additional file 3: Text S1.

In addition to using LLMs and existing similarity evaluation tools (i.e., AST and ROUGE) to assess the code, we also conducted a lightweight execution-grounded check from data preprocessing outputs and two manually defined checks. For data preprocessing outputs, we extracted the data preprocessing code snippet, executed it, and verified the number of cells and genes in the processed data. If the results exactly matched those from the reference code, the implementation was considered consistent. Otherwise, it was marked as inconsistent. The consistency rate during the preprocessing stage is presented in Additional file 1: Fig. S3b. For two manually defined checks, we selected 13 tasks from 8 representative task categories as an evaluation subset (Additional file 1: Fig. S3c). For these checks, we performed code separation and verification by combining manual review with a LLM. We used a binary 0-1 scoring system. For numerical verification, a score of 1 was assigned for matching the ground truth and 0 for not matching. For validity judgment verification, a score of 1 was given if the judgment criteria were met, and 0 otherwise. Additional file 1: Fig. S3e provided detailed execution-grounded check results. All these results serve as a supplementary evaluation of the code. More details are provided in Additional file 3: Text S2.

**Experimental environment.** Our computational infrastructure comprised a dedicated server housing an NVIDIA A100 GPU with 80 GB of high-bandwidth memory (VRAM) and a multi-core CPU (32 cores). Furthermore, in the evaluation pipeline, we also employed one BW1000 (64G) intelligent acceleration card—domestically produced in China and provided by Dawning Information Industry Co.,Ltd. To address potential dependency conflicts arising from the diverse software requirements of the 50 single-cell omics analysis tasks, we meticulously configured and utilized five separate Python virtual environments (managed with conda) and three independent R environments (managed with r-conda). This deliberate isolation of computational environments was critical for maintaining version consistency, avoiding library conflicts, and ensuring the reliable and reproducible execution of all benchmark experiments.

Key computational resources leveraged in this study include established bioinformatics toolkits such as scvi-tools, scanpy and scarches, alongside widely-used standalone packages like spage, cellphonedb and mira-multiome. A comprehensive listing of all software packages employed, including their specific versions and the designated environment for experimental tasks, is provided in Additional file 1: Table S10. We have provided all isolated Conda environments on Zenodo link at https://doi.org/10.5281/zenodo.17455069 to ensure the reproducibility of our evaluation benchmark for bioinformatics agents.

## Availability of data and materials

All datasets used in the benchmark are published at the specific Zendo link: https://zenodo.org/records/17291196 [93]. Datasets, ground-truth scripts/notebooks, groundtruth results, database for KB construction, input prompts for all 50 single-cell omics analysis tasks and multiple-datasets tasks have been contained in the zip file. Please see README.md in the file or the code link for more detailed information. The workflow and evaluation pipeline in this study is freely avaliable under the MIT license at https://github.com/lyyang01/bioagent-benchmark [94] , and these codes are also deposited at Zendo: https://doi.org/10.5281/zenodo.18437898 [95].

## Funding

This work was supported by Major Project of Guangzhou National Laboratory (No. SRPG22-007); National Key R&D Program of China (2023YFF1204701);

National Natural Science Foundation of China (No. 12371485 and No. 82400622); Prevention and Control of Emerging and Major Infectious Diseases-National Science and Technology Major Project (2025ZD01901900); Youth science foundation of Guangzhou National Laboratory (grant no. QNPG23-12); Major Project of Guangzhou National Laboratory GZNL2025C01013; Guangdong Basic and Applied Basic Research Foundation (No.2025A1515011597);

## Acknowledgements

We acknowledge technical support from Data Science Platform of Guangzhou National Laboratory and Bio-medical big data Operating System (Bio-OS).

## Peer review information

## Authors' contributions

Conceptualization, R.S. and Y.L.; Supervision, R.H., X.Z., R.S. and Y.L.; Datasets collection, processing, and application, R.S., Y.L., and L.Z.; Experimental implementation, Y.L. and L.Z.; Methods comparisons and analysis, Y.L., L.Z., and R.S.; Manuscript writing and figure generation, Y.L., L.Z., R.S. and X.D.; Manuscript reviewing, R.S. and Y.L. All authors read and approved the final manuscript.

## Ethics approval and consent to participate

Not applicable

# Consent for publication

Not applicable

# Competing interests

X.D. is employed by Dawning Information Industry (Beijing) Co.,Ltd, which provided BW1000 (64G) intelligent acceleration cards for this study, and the author declares no other competing interests. X.Z. is employed by Feihe Reseach Institute of Heilongjiang Feihe Dairy Co., Ltd. and the author declares that this employment has no direct or relevant competing interests in relation to the specific topic, content, and conclusions of this study. The remaining authors declare no competing interests.

# Supplementary Information

Additional File 1: Supplementary Figures. This file contains Figs. S1 to S13. Fig. S1: The features of current agent benchmarks in bioinformatics, LLMs and agent frameworks. Fig. S2: Technical details of the main experiment. Figs. S3–S5: Supplementary results of the main experiment. Figs. S6–S7: Association between result consistency and biological matrices. Fig. S8: Prompt templates. Figs. S9–S12: Agent robustness, functional modules, and analysis of failed tasks. Fig. S13: Parameters in the evaluation metrics.

Additional File 2: Supplementary Tables. This file contains Tables S1 to S13. Table S1-S2: Detailed task information. Table S3, S5, S7-S8: Prompt for workflow, evaluation and ablation experiments. Table S4: Result consistency for individual task. Table S9: Error types and definitions. Table S10:

Environment. Table S11: Costs. Table S12: Token lengths on the prompt-tier setting. Table S13: Sensitivity analysis.

Additional File 3: Supplementary Texts. This file contains Text S1 to S3. Text S1: Comparisons of single-judge scoring and cross-judge scoring with LLMs. Text S2: Execution-grounded checks for the agent-generated codes. Text S3: Complete titles of the supplementary tables.

# Reference

1. Vandereyken, Katy et al. "Methods and applications for single-cell and spatial multi-omics." Nature reviews. Genetics vol. 24,8 (2023): 494-515. doi:10.1038/s41576-023-00580-2
2. Regev, Aviv et al. "The Human Cell Atlas." eLife vol. 6 e27041. 5 Dec. 2017, doi:10.7554/eLife.27041
3. Tian, Luyi et al. "Benchmarking single cell RNA-sequencing analysis pipelines using mixture control experiments." Nature methods vol. 16,6 (2019): 479-487. doi:10.1038/s41592-019-0425-8
4. Marcondes, D., et al. "Back to Basics to Open the Black Box." Nature Machine Intelligence, vol. 6 (2024): 498–501. doi:10.1038/s42256-024-00842-6.
5. Pliner, Hannah A et al. "Supervised classification enables rapid annotation of cell atlases." Nature methods vol. 16,10 (2019): 983-986. doi:10.1038/s41592-019-0535-3
6. Cao, Zhi-Jie, and Ge Gao. "Multi-omics single-cell data integration and regulatory inference with graph-linked embedding." Nature biotechnology vol. 40,10 (2022): 1458-1466. doi:10.1038/s41587-022-01284-4
7. Wei, Jason, et al. "Chain-of-Thought Prompting Elicits Reasoning in Large Language Models." Advances in Neural Information Processing Systems, vol. 35 (2022): 24824–24837

8. Park, Joon Sung, et al. "Generative Agents: Interactive Simulacra of Human Behavior." Proceedings of the 36th Annual ACM Symposium on User Interface Software and Technology, ACM, 2023, pp. 1–22, doi:10.1145/3586183.3606763.
9. Lewis, Patrick, et al. "Retrieval-Augmented Generation for Knowledge-Intensive NLP Tasks." Advances in Neural Information Processing Systems, vol. 33 (20220): 9459–9474.
10. Wang, Hanchen, et al. "SpatialAgent: An autonomous AI agent for spatial biology." *bioRxiv* (2025): 2025-04.
11. Huang, Kexin, et al. "Biomni: A general-purpose biomedical ai agent." *Biorxiv* (2025).
12. Alber, Samuel, et al. "Cellvoyager: Ai compbio agent generates new insights by autonomously analyzing biological data." bioRxiv (2025): 2025-06.
13. Phan, Long, et al. "Humanity's last exam." *arXiv preprint arXiv:2501.14249* (2025).
14. Mangul, Serghei, et al. "BioLLMBench: A Comprehensive Benchmarking of Large Language Models in Bioinformatics." bioRxiv, preprint, 19 Dec. 2023, https://doi.org/10.1101/2023.12.19.572483.
15. Mitchener, Robert, et al. "BixBench: A Comprehensive Benchmark for LLM-based Agents in Computational Biology." arXiv, preprint, arXiv:2504.01986v2, 2025. https://arxiv.org/abs/2504.01986.
16. Luo, Yifei, et al. "Benchmarking AI Scientists in Omics Data-Driven Biological Research." arXiv, preprint, arXiv:2505.08341v1, 2025. https://arxiv.org/abs/2505.08341.
17. Wang, Chen, et al. "GenoTEX: A Benchmark for Evaluating LLM-Based Exploration of Gene Expression Data in Alignment with Bioinformaticians." arXiv, preprint, arXiv:2406.15341v3, 2024. https://arxiv.org/abs/2406.15341.
18. Tang, Xiangru et al. "BioCoder: a benchmark for bioinformatics code generation with large language models." Bioinformatics (Oxford, England) vol. 40,Suppl 1 (2024): i266-i276. doi:10.1093/bioinformatics/btae230

19. Laurent, Jon M., et al. "Lab-bench: Measuring Capabilities of Language Models for Biology Research." arXiv, preprint, arXiv:2407.10362, 2024. arxiv.org/abs/2407.10362.
20. Chen, Ziru, et al. "Scienceagentbench: Toward Rigorous Assessment of Language Agents for Data-Driven Scientific Discovery." arXiv, preprint, arXiv:2410.05080, 2024. arxiv.org/abs/2410.05080.
21. Yao, Shunyu, et al. "ReAct: Synergizing Reasoning and Acting in Language Models." Proceedings of the 11th International Conference on Learning Representations. 2023.
22. LangChain Inc. "LangGraph: Multi-Agent Workflows." LangChain Documentation, version 0.1.0, 2024. python.langchain.com/v0.1/docs/langgraph/.
23. Wu, Qingyun, et al. "AutoGen: Enabling Next-Gen LLM Applications via Multi-Agent Conversation Framework." arXiv, preprint, arXiv:2308.08155, 2023. arxiv.org/abs/2308.08155.
24. OpenAI. GPT-4o System Card. arXiv, 25 Oct. 2024, https://arxiv.org/abs/2410.21276.
25. OpenAI. "Introducing GPT-4.1 in the API." OpenAI, 14 Apr. 2025, https://openai.com/index/gpt-4-1/.
26. DeepSeek-AI, et al. DeepSeek-R1: Incentivizing Reasoning Capability in LLMs via Reinforcement Learning. arXiv, 22 Jan. 2025, https://arxiv.org/abs/2501.12948.
27. DeepSeek-AI, et al. DeepSeek-V3 Technical Report. arXiv, 18 Feb. 2025, https://arxiv.org/abs/2412.19437.
28. Qwen Team. Qwen2.5 Technical Report. arXiv, 3 Jan. 2025, https://doi.org/10.48550/arXiv.2412.15115.
29. Anthropic. Claude 3.7 Sonnet System Card. Feb. 2025, https://assets.anthropic.com/m/785e231869ea8b3b/original/claude-3-7-sonnet-system-card.pdf

30. Google, Gemini Team. Gemini 2.5: Pushing the Frontier with Advanced Reasoning, Multimodality, Long Context, and Next Generation Agentic Capabilities. 17 June 2025, https://storage.googleapis.com/deepmind-media/gemini/gemini_v2_5_report.pdf.
31. xAI. "Grok 3 Beta — The Age of Reasoning Agents." xAI, 19 Feb. 2025, https://x.ai/news/grok-3.
32. Liu, Nelson F., et al. "Lost in the Middle: How Language Models Use Long Contexts." *arXiv*, 6 July 2023, doi:10.48550/arXiv.2307.03172.
33. Gao, Muhan, et al. "Insights into LLM Long-Context Failures: When Transformers Know but Don't Tell." *Findings of the Association for Computational Linguistics: EMNLP 2024*, 2024, pp. 7611–25, doi:10.18653/v1/2024.findings-emnlp.447.
34. Richens, Jonathan, et al. General Agents Need World Models. Proceedings of the 42nd International Conference on Machine Learning (ICML), vol. 267, PMLR, 2025, Vancouver, Canada.
35. Langchain-ai. Langchain. GitHub, https://github.com/langchain-ai/langchain.
36. Chroma-core. Chroma. GitHub, https://github.com/chroma-core/chroma.
37. Yang, Wuu. "Identifying syntactic differences between two programs." Software: Practice and Experience 21.7 (1991): 739-755.
38. Lin, Chin-Yew. "Rouge: A package for automatic evaluation of summaries." Text summarization branches out. 2004.
39. Lavie, Alon, and Michael J. Denkowski. "The METEOR metric for automatic evaluation of machine translation." Machine translation 23 (2009): 105-115.
40. Chaudhuri, Surajit, et al. "Robust and efficient fuzzy match for online data cleaning." Proceedings of the 2003 ACM SIGMOD international conference on Management of data. 2003.
41. Chen, Mark, et al. "Evaluating large language models trained on code." arXiv preprint arXiv:2107.03374 (2021).

42. Santini, Simone, and Ramesh Jain. "Similarity measures." IEEE Transactions on pattern analysis and machine Intelligence 21.9 (1999): 871-883.

43. Hotelling, Harold. "Analysis of a complex of statistical variables into principal components." Journal of educational psychology 24.6 (1933): 417.

44. Lopez, Romain et al. "Deep generative modeling for single-cell transcriptomics." Nature methods vol. 15,12 (2018): 1053-1058. doi:10.1038/s41592-018-0229-2

45. Xu, Chenling et al. "Probabilistic harmonization and annotation of single-cell transcriptomics data with deep generative models." Molecular systems biology vol. 17,1 (2021): e9620. doi:10.15252/msb.20209620

46. Hie, Brian et al. "Efficient integration of heterogeneous single-cell transcriptomes using Scanorama." Nature biotechnology vol. 37,6 (2019): 685-691. doi:10.1038/s41587-019-0113-3

47. Korsunsky, Ilya et al. "Fast, sensitive and accurate integration of single-cell data with Harmony." Nature methods vol. 16,12 (2019): 1289-1296. doi:10.1038/s41592-019-0619-0

48. De Donno, Carlo et al. "Population-level integration of single-cell datasets enables multi-scale analysis across samples." Nature methods vol. 20,11 (2023): 1683-1692. doi:10.1038/s41592-023-02035-2

49. Lotfollahi, Mohammad et al. "scGen predicts single-cell perturbation responses." Nature methods vol. 16,8 (2019): 715-721. doi:10.1038/s41592-019-0494-8

50. Zhang, Allen W et al. "Probabilistic cell-type assignment of single-cell RNA-seq for tumor microenvironment profiling." Nature methods vol. 16,10 (2019): 1007-1015. doi:10.1038/s41592-019-0529-1

51. Domínguez Conde, C et al. "Cross-tissue immune cell analysis reveals tissue-specific features in humans." Science (New York, N.Y.) vol. 376,6594 (2022): eabl5197. doi:10.1126/science.abl5197

52. Badia-I-Mompel, Pau et al. “decoupleR: ensemble of computational methods to infer biological activities from omics data.” Bioinformatics advances vol. 2,1 vbac016. 8 Mar. 2022, doi:10.1093/bioadv/vbac016

53. Wolf, F Alexander et al. “PAGA: graph abstraction reconciles clustering with trajectory inference through a topology preserving map of single cells.” Genome biology vol. 20,1 59. 19 Mar. 2019, doi:10.1186/s13059-019-1663-x

54. Bergen, Volker et al. “Generalizing RNA velocity to transient cell states through dynamical modeling.” Nature biotechnology vol. 38,12 (2020): 1408-1414. doi:10.1038/s41587-020-0591-3

55. Lange, Marius et al. “CellRank for directed single-cell fate mapping.” Nature methods vol. 19,2 (2022): 159-170. doi:10.1038/s41592-021-01346-6

56. Cannoodt, Robrecht, et al. "SCORPIUS Improves Trajectory Inference and Identifies Novel Modules in Dendritic Cell Development." bioRxiv preprint, 8 Nov. 2016, https://doi.org/10.1101/079509.

57. Weinberger, Ethan et al. “Isolating salient variations of interest in single-cell data with contrastiveVI.” Nature methods vol. 20,9 (2023): 1336-1345. doi:10.1038/s41592-023-01955-3

58. Ashuach, Tal et al. “PeakVI: A deep generative model for single-cell chromatin accessibility analysis.” Cell reports methods vol. 2,3 100182. 15 Mar. 2022, doi:10.1016/j.crmeth.2022.100182

59. Martens, Laura D et al. “Modeling fragment counts improves single-cell ATAC-seq analysis.” Nature methods vol. 21,1 (2024): 28-31. doi:10.1038/s41592-023-02112-6

60. Gayoso, Adam et al. “Joint probabilistic modeling of single-cell multi-omic data with totalVI.” Nature methods vol. 18,3 (2021): 272-282. doi:10.1038/s41592-020-01050-x

61. Ashuach, Tal et al. "MultiVI: deep generative model for the integration of multimodal data." Nature methods vol. 20,8 (2023): 1222-1231. doi:10.1038/s41592-023-01909-9
62. Lynch, Allen W et al. "MIRA: joint regulatory modeling of multimodal expression and chromatin accessibility in single cells." Nature methods vol. 19,9 (2022): 1097-1108. doi:10.1038/s41592-022-01595-z
63. Argelaguet, Ricard et al. "MOFA+: a statistical framework for comprehensive integration of multi-modal single-cell data." Genome biology vol. 21,1 111. 11 May. 2020, doi:10.1186/s13059-020-02015-1
64. Hao, Yuhan et al. "Dictionary learning for integrative, multimodal and scalable single-cell analysis." Nature biotechnology vol. 42,2 (2024): 293-304. doi:10.1038/s41587-023-01767-y
65. Lopez, Romain et al. "DestVI identifies continuums of cell types in spatial transcriptomics data." Nature biotechnology vol. 40,9 (2022): 1360-1369. doi:10.1038/s41587-022-01272-8
66. Andersson, Alma et al. "Single-cell and spatial transcriptomics enables probabilistic inference of cell type topography." Communications biology vol. 3,1 565. 9 Oct. 2020, doi:10.1038/s42003-020-01247-y
67. Kleshchevnikov, Vitalii et al. "Cell2location maps fine-grained cell types in spatial transcriptomics." Nature biotechnology vol. 40,5 (2022): 661-671. doi:10.1038/s41587-021-01139-4
68. Biancalani, Tommaso et al. "Deep learning and alignment of spatially resolved single-cell transcriptomes with Tangram." Nature methods vol. 18,11 (2021): 1352-1362. doi:10.1038/s41592-021-01264-7
69. Long, Yahui et al. "Spatially informed clustering, integration, and deconvolution of spatial transcriptomics with GraphST." Nature communications vol. 14,1 1155. 1 Mar. 2023, doi:10.1038/s41467-023-36796-3

70. Moriel, Noa et al. “NovoSpaRc: flexible spatial reconstruction of single-cell gene expression with optimal transport.” Nature protocols vol. 16,9 (2021): 4177-4200. doi:10.1038/s41596-021-00573-7

71. Lopez, Romain, et al. "A Joint Model of Unpaired Data from scRNA-seq and Spatial Transcriptomics for Imputing Missing Gene Expression Measurements." arXiv, Cornell University, 6 May 2019, arxiv.org/abs/1905.02269.

72. Abdelaal, Tamim et al. “SpaGE: Spatial Gene Enhancement using scRNA-seq.” Nucleic acids research vol. 48,18 (2020): e107. doi:10.1093/nar/gkaa740

73. Xu, Hang et al. “Unsupervised spatially embedded deep representation of spatial transcriptomics.” Genome medicine vol. 16,1 12. 12 Jan. 2024, doi:10.1186/s13073-024-01283-x

74. Hu, Jian et al. “SpaGCN: Integrating gene expression, spatial location and histology to identify spatial domains and spatially variable genes by graph convolutional network.” Nature methods vol. 18,11 (2021): 1342-1351. doi:10.1038/s41592-021-01255-8

75. Palla, Giovanni et al. “Squidpy: a scalable framework for spatial omics analysis.” Nature methods vol. 19,2 (2022): 171-178. doi:10.1038/s41592-021-01358-2

76. DeTomaso, David, and Nir Yosef. “Hotspot identifies informative gene modules across modalities of single-cell genomics.” Cell systems vol. 12,5 (2021): 446-456.e9. doi:10.1016/j.cels.2021.04.005

77. Garcia-Alonso, Luz et al. “Single-cell roadmap of human gonadal development.” Nature vol. 607,7919 (2022): 540-547. doi:10.1038/s41586-022-04918-4

78. Browaeys, Robin et al. “NicheNet: modeling intercellular communication by linking ligands to target genes.” Nature methods vol. 17,2 (2020): 159-162. doi:10.1038/s41592-019-0667-5

79. Jin, Suoqin et al. “Inference and analysis of cell-cell communication using CellChat.” Nature communications vol. 12,1 1088. 17 Feb. 2021, doi:10.1038/s41467-021-21246-9

80. Traag, V A et al. "From Louvain to Leiden: guaranteeing well-connected communities." Scientific reports vol. 9,1 5233. 26 Mar. 2019, doi:10.1038/s41598-019-41695-z

81. He, Peng et al. "A human fetal lung cell atlas uncovers proximal-distal gradients of differentiation and key regulators of epithelial fates." Cell vol. 185,25 (2022): 4841-4860.e25. doi:10.1016/j.cell.2022.11.005

82. He, Peng et al. "The changing mouse embryo transcriptome at whole tissue and single-cell resolution." Nature vol. 583,7818 (2020): 760-767. doi:10.1038/s41586-020-2536-x

83. Zhang, Ziqi et al. "scMoMaT jointly performs single cell mosaic integration and multi-modal bio-marker detection." Nature communications vol. 14,1 384. 24 Jan. 2023, doi:10.1038/s41467-023-36066-2

84. M Ascensión, Alex et al. "Triku: a feature selection method based on nearest neighbors for single-cell data." GigaScience vol. 11 (2022): giac017. doi:10.1093/gigascience/giac017

85. Vandenbon, Alexis, and Diego Diez. "A clustering-independent method for finding differentially expressed genes in single-cell transcriptome data." Nature communications vol. 11,1 4318. 28 Aug. 2020, doi:10.1038/s41467-020-17900-3

86. Song, Dongyuan et al. "scPNMF: sparse gene encoding of single cells to facilitate gene selection for targeted gene profiling." Bioinformatics (Oxford, England) vol. 37,Suppl_1 (2021): i358-i366. doi:10.1093/bioinformatics/btab273

87. Townes, F William et al. "Feature selection and dimension reduction for single-cell RNA-Seq based on a multinomial model." Genome biology vol. 20,1 295. 23 Dec. 2019, doi:10.1186/s13059-019-1861-6

88. Mathys, Hansruedi et al. "Single-cell multiregion dissection of Alzheimer's disease." Nature vol. 632,8026 (2024): 858-868. doi:10.1038/s41586-024-07606-7

89. Love, Michael I et al. “Moderated estimation of fold change and dispersion for RNA-seq data with DESeq2.” Genome biology vol. 15,12 (2014): 550. doi:10.1186/s13059-014-0550-8

90. Choi, Hyunsu, et al. "ResolVI - Addressing Noise and Bias in Spatial Transcriptomics." bioRxiv preprint, 20 Jan. 2025, https://doi.org/10.1101/2025.01.20.634005.

91. Lause, Jan et al. “Analytic Pearson residuals for normalization of single-cell RNA-seq UMI data.” Genome biology vol. 22,1 258. 6 Sep. 2021, doi:10.1186/s13059-021-02451-7

92. Cemri, Mert, et al. Why Do Multi-Agent LLM Systems Fail? arXiv, 22 Apr. 2025, arXiv:2503.13657v2 [cs.AI]. https://arxiv.org/abs/2503.13657.

93. Yang Liu, Lu Zhou, Xiawei Du, Ruikun He, Xuguang Zhang, Rongbo Shen, Yixue Li. Bechmarking LLM-based agents for single-cell omics analysis. Datasets. Zeonodo. https://doi.org/10.5281/zenodo.17291196 (2025).

94. Yang Liu, Lu Zhou, Xiawei Du, Ruikun He, Xuguang Zhang, Rongbo Shen, Yixue Li. Bechmarking LLM-based agents for single-cell omics analysi. Github. https://github.com/lyyang01/bioagent-benchmark (2025).

95. Yang Liu, Lu Zhou, Xiawei Du, Ruikun He, Xuguang Zhang, Rongbo Shen, Yixue Li. Bechmarking LLM-based agents for single-cell omics analysi. Zenodo. https://doi.org/10.5281/zenodo.18437898 (2025).

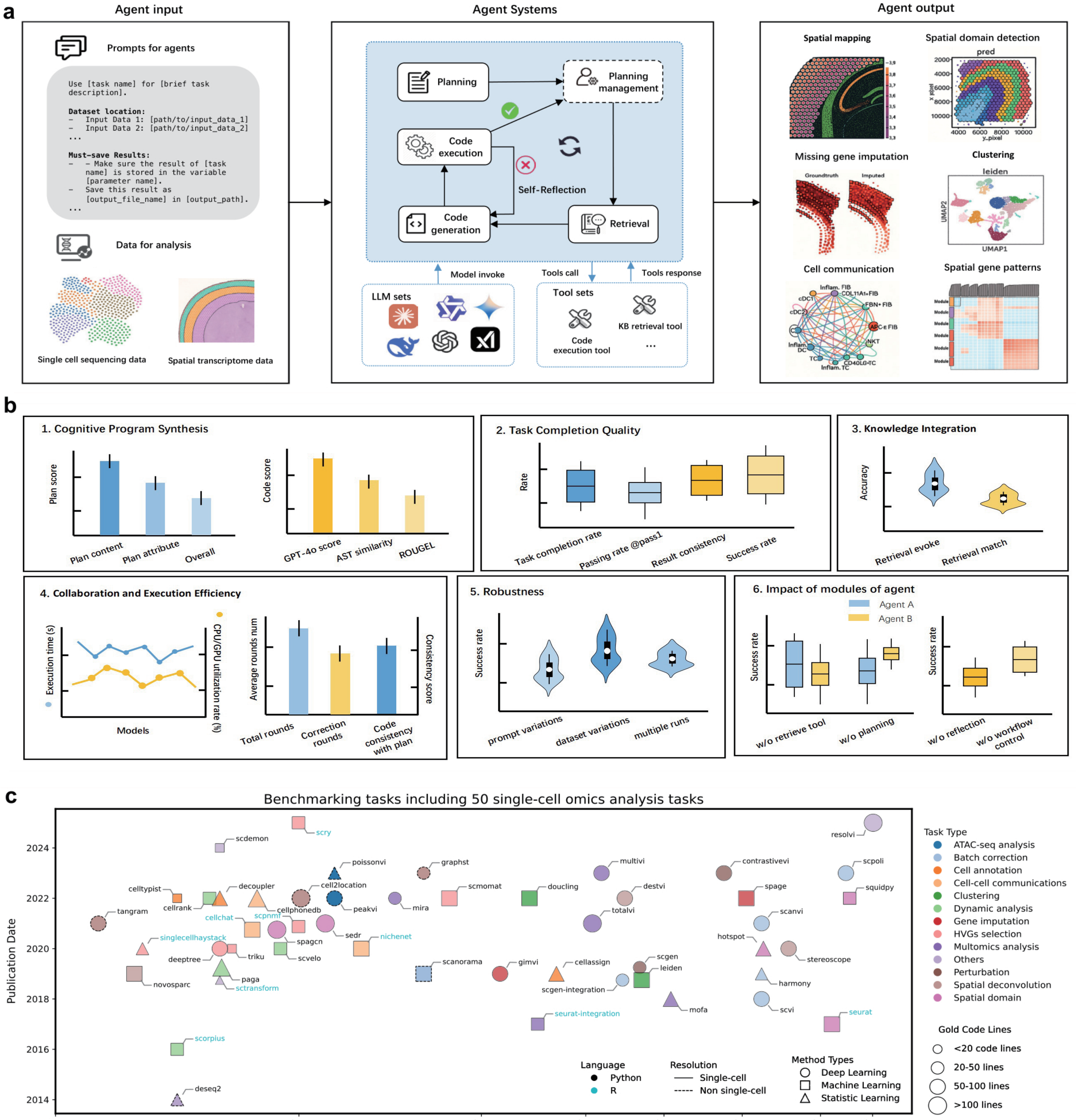

**Figure 1. Benchmarking evaluation system. a. Benchmarking platform.** Agent systems take two types of input: task-specific prompts and raw biological data. The systems integrate multiple LLMs, a suite of auxiliary tools, and support various agent frameworks. The output consists of visualizations and analysis results. **b. Evaluation metrics.** A set of 18 evaluation metrics was developed to assess four major dimensions: cognitive program synthesis, task completion quality, knowledge integration, and collaboration and execution efficiency. In addition, the robustness and the impact of functional modules were further investigated. **c. Benchmarking tasks.** 50 representative single-cell omics analysis tasks were compiled to perform systematic evaluation, spanning multi-omics, multiple species, and various sequencing technologies.

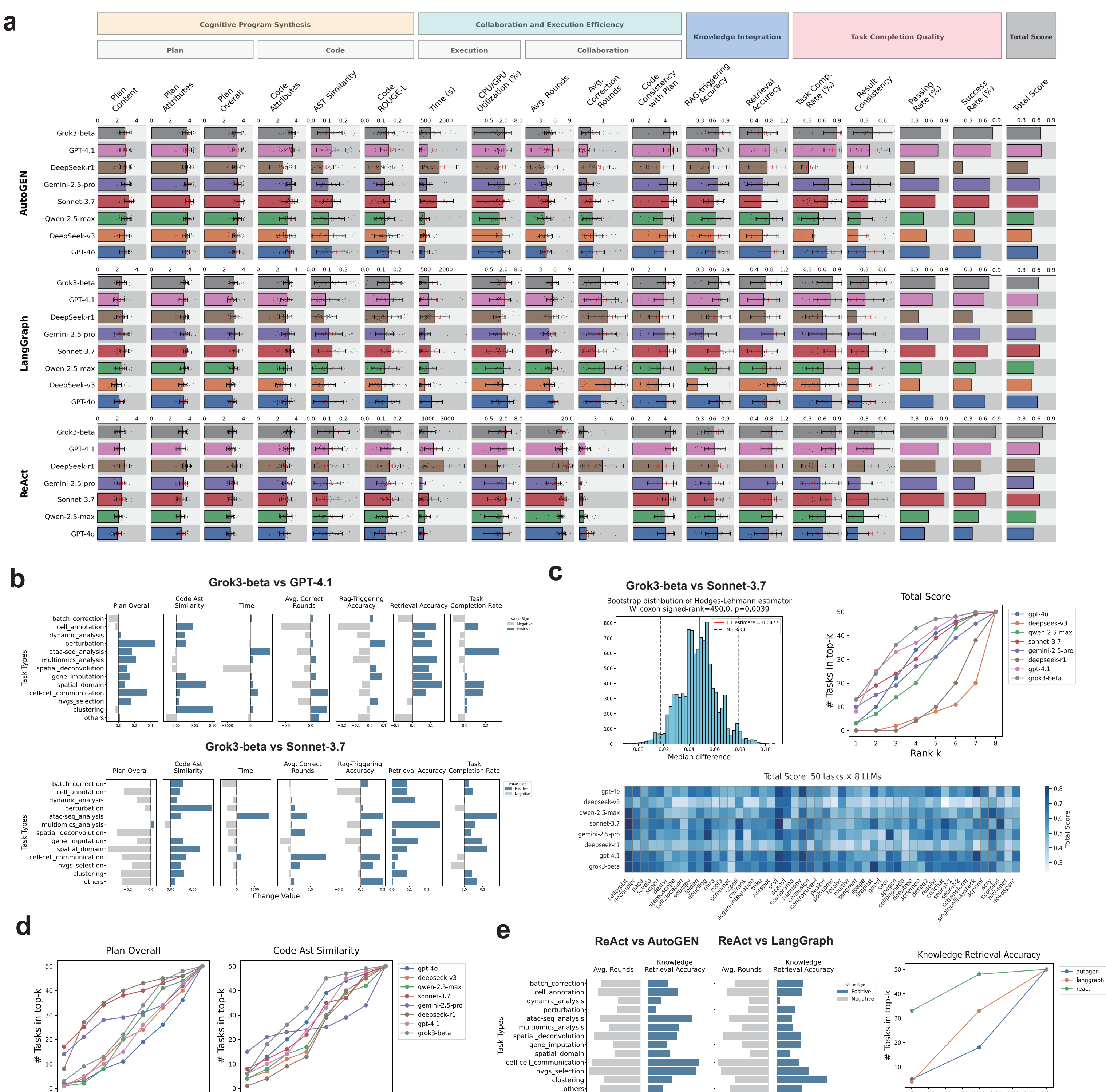

**Figure 2. a. Benchmarking results of multiple agent frameworks cross diverse LLMs.** Three agent frameworks (i.e., AutoGEN, LangGraph, ReAct) combined with eight widely used LLMs (i.e., GPT-4o, GPT-4.1, DeepSeek-R1, DeepSeek-V3, Qwen-2.5-max, Sonnet-3.7, Gemini-2.5-pro, Grok3-beta) were evaluated on benchmarking tasks. The final column shows the Total Score (0–1, higher = better), aggregated from the first 17 metrics. Individual measurements from 50 benchmarking tasks per agent framework-LLM combination are depicted as scatter points, error bars represent ±1 standard deviation (black) and 95% CI (red) of these measurements. **b. Grok3-beta vs other LLMs.** Differences between Grok3-beta and GPT-4.1/Sonnet-3.7 across specific metric, categorized by task types. **c. Total score analysis for 50 tasks.** Comparison of total scores for each task across 8 LLMs. Bootstrap distribution of HL estimate on Grok3-beta and Sonnet-3.7 is also shown in the figure. **d. Count of how many times each LLM lands in the top-k across 50 tasks. e. ReAct vs other agent frameworks.** Avg. Rounds and Knowledge Retrieval Accuracy by ReAct, AutoGEN and LangGraph are compared to show the high retrieval accuracy yet severe round redundancy of the single-agent framework.

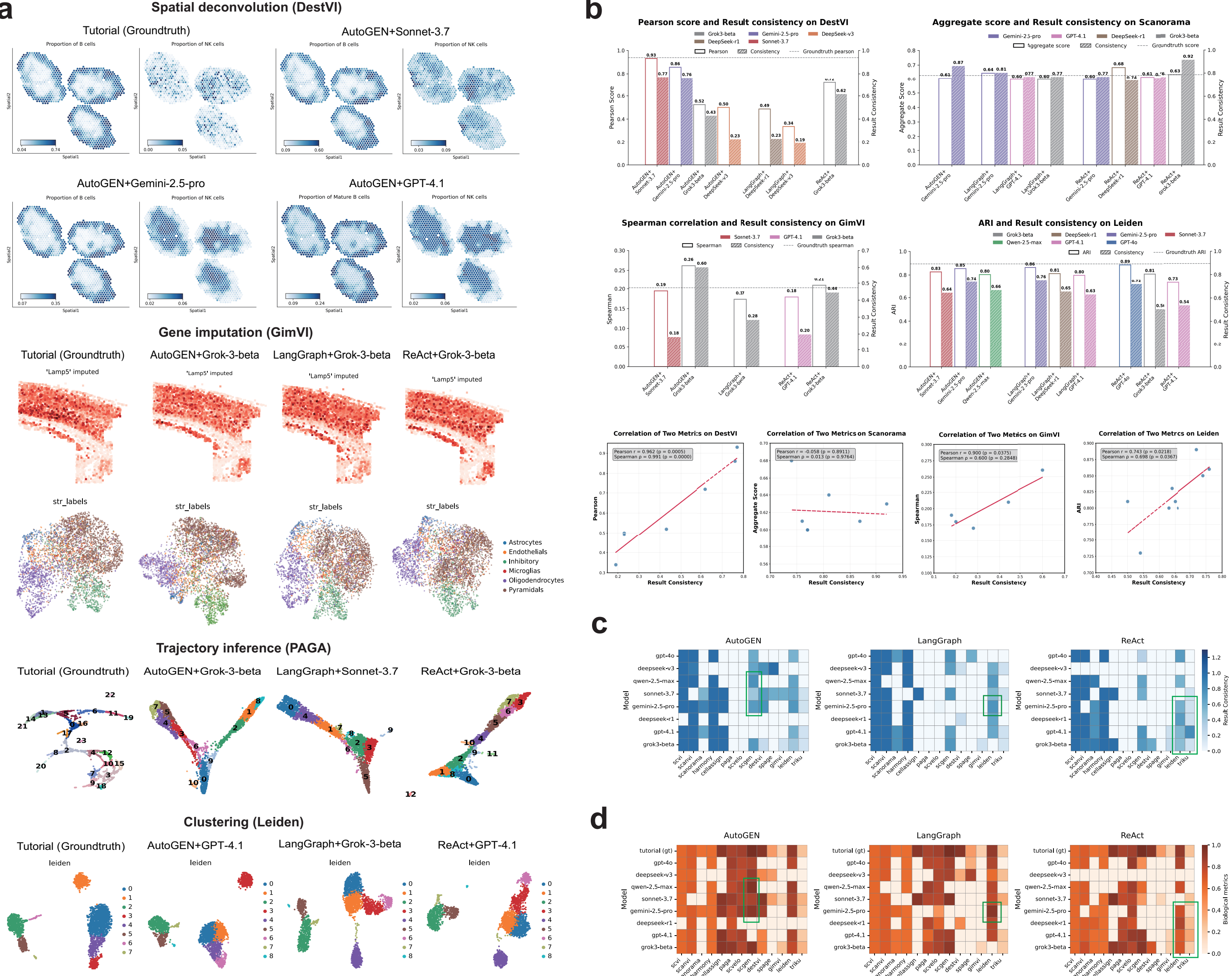

**Figure 3. Analysis on result consistency and biological metric. a. Visualization of biological results.** Four tasks—spatial deconvolution, gene imputation, trajectory inference, and clustering—were selected to show the results produced by the groundtruth script (from tutorial) and the results produced by the agents. **b. Result consistency vs. biological metrics.** Performance is shown for four tasks: DestVI, Scanorama, GimVI, and Leiden. The left axis represents the biological metric, displayed as unfilled bars; the right axis indicates results consistency, shown as shaded bars. A horizontal dashed line marks the biological metric value obtained by the groundtruth script. Correlations between results consistency and the biological metric across the four tasks are also shown. **c. Result consistency on a subset task.** They were grouped by three agent frameworks: AutoGEN, LangGraph and ReAct. **d. Biological metrics on a subset task.** They were grouped by three agent frameworks: AutoGEN, LangGraph and ReAct. Note that “tutorial (gt)” represents the biological metric value computed from results obtained by the groundtruth script.

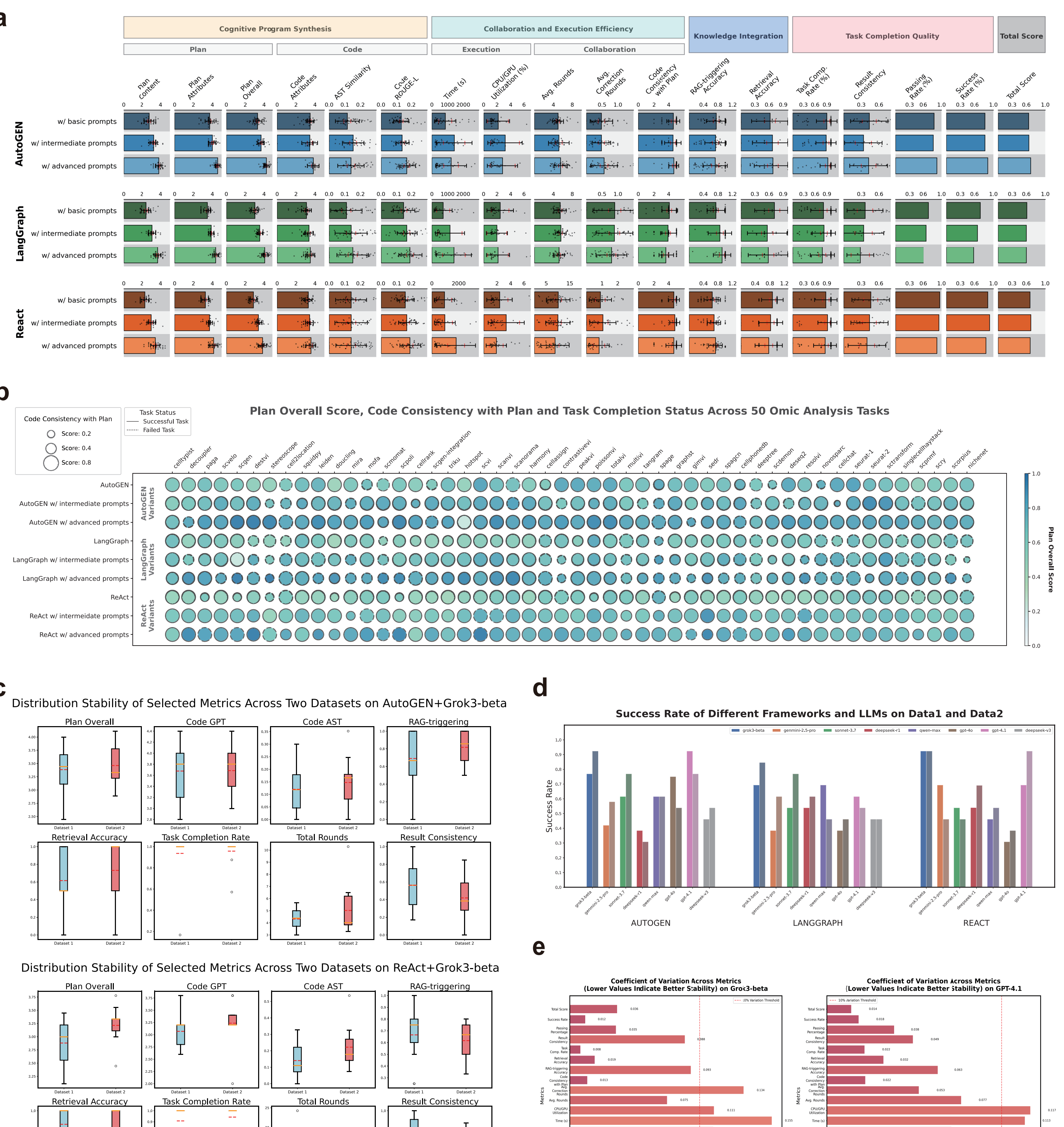

**Figure 4. Assessing the robustness of prompt variations, dataset variations and multiple runs. a. Results of multiple agent frameworks with basic-, intermediate-, advanced- prompts.** Performance comparison of three prompt tiers including Basic, Intermediate, and Advanced evaluated using the top-performing model Grok3-beta. **b. Relationship of task success to plan overall score and code consistency with plan.** Bubble size represents code consistency with plan (bigger sizes represent higher scores) while color intensity represents plan overall score (deeper colors represent higher scores). **c. Distribution stability of specific metrics across two datasets of same task types.** Eight metrics are selected to show the comparison of two distinct sets of datasets across frameworks and LLMs. The boxplots display the median (orange line), mean (red dashed line), and the range of the data within 1.5 times the IQR (whiskers). **d. Comparison of success rate across agent frameworks and LLMs on the two distinct set of datasets.** These datasets are simplified noted as Data1 and Data2. Each set of datasets contains 13 datasets for the same 13 tasks, which is shown in the Supplementary Fig 10a. **e. Variation across metrics of multiple runs on different LLMs.** The red dashed line in the figure represents the 10% variation boundary.

**a**

**b**

Code Quaility, Average Rounds and Task Completion Status Across 50 Omic Analysis Tasks

**Figure 5. Assessing the impact of functional modules. a. The results of functional module ablations. Systematic functional module analysis is performed on the ReAct and AutoGEN frameworks (using Grok-3-beta).** Four functional modules are evaluated: knowledge retrieval, planning, reflection, and inter-agent workflow control. ReAct is ablated only for knowledge retrieval and planning, because ReAct inherently lacks removable reflection modules and inter-agent communication mechanisms. **b. Task success vs. code quality and average rounds.** Bubble size corresponds to code quality (synthesized from code attribute, code AST similarity, and code ROUGE-L scores). Color intensity signifies the average number of collaboration rounds (darker = fewer collaboration rounds, normalized to [0, 1]). Line type denotes task completion status (solid for success, dashed for failure).

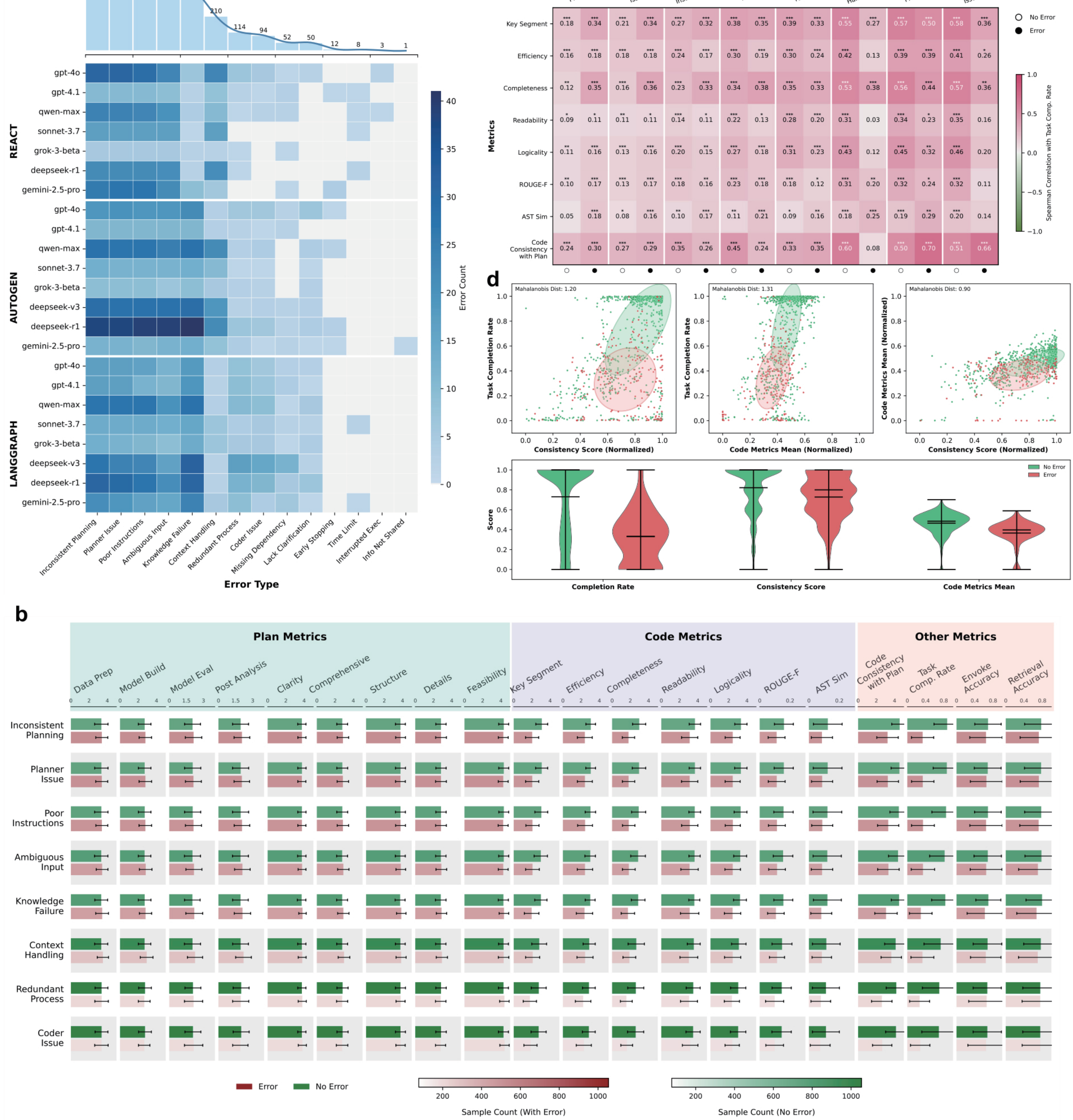

**Figure 6. Analysis of failed tasks. a. Distribution of error types across agent frameworks and LLMs.** Heatmap shows the distribution of error types observed during task execution across three agent frameworks and eight LLMs. Inconsistent planning behavior, planner issues, and poor instruction following were the most frequently observed. **b. Association between key error types and fine-grained performance metrics.** For the eight most frequent error types, significant differences in code quality, code consistency with plan, and task completion rate, suggest a measurable performance degradation associated with these errors. **c.** Correlation structure shifts in error-present vs. error-absent groups. Heatma shows Spearman correlation patterns between task completion rate and key performance metrics (i.e., code metrics, code consistency with plan) for error and no-error groups. **d. Case analysis of long context handling failure.** A detailed comparison between error(red) and no-error(green) groups for this error type shows that the no-error group consistently outperformed the error group in task completion rate, code consistency with plan, and other code metrics.

**a**

| | Evaluation Metrics | | Task Coverage | | | Platform Universality | | | | Core Capability Insights | Strengths |
|---|---|---|---|---|---|---|---|---|---|---|---|
| | # of Metrics | Dimensions | Task form | # of Scenarios | Scenarios Types | Multi -Lang. | Open-Souce | # of test LLM | # of test agent frameworks | | |
| BioLLMBench (2024) | 8 | Quality of code and plan | QA | 36 | Conceptual QA, code completion, visual description | ✗ | ✗ | 3 | - | ❌Only reports overall CPI and single-task accuracy | ▶ Bioinformatics benchmark with fine-grained manual evaluation |
| Lab-bench (2024) | 3 | Task accuracy, task precision and human evaluation | QA | 8 | Scenarios related to multiple plasmids and DNA fragments | ✗ | ✓ | 7 | - | ❌Lacks quantification of module impact | ▶ Cross-disciplinary coverage |
| BaisBench (2024) | 1 | Task-specific accuracy (e.g., cell annotation score) | Modular tasks | 2 | Cell annotation | ✗ | ✗ | 2 | - | ❌Merely reports failure rates | ▶ Emphasizes downstream reasoning |
| GenoTEX (2025) | 8 | GTA analysis, trait prediction and runtime | Modular tasks | 3 | Gene expression data filtering, preprocess and analyisis | ✗ | ✗ | 1 | - | ❌Over-simplifies attribution to model knowledge gaps | ▶ Extensive gene expression analysis tasks |
| BixBench (2025) | 2 | open-answer and multi-choice accuracy | QA and multi-choice | 53 | Literature mining, DNA engineering | ✓ | ✓ | 2 | - | ❌Fails to analyze module contribution or failure cause | ▶ General bioinformatics scenarios |
| BioCoder (2025) | 1 | Code execution pass rate | Bioinformatics functions | 8 | Cross-disciplinary applications | ✓ | ✓ | 8 | - | ❌Only tracks superficial errors (syntax/output mismatch) | ▶ Large-scale code generation evaluation |
| ScienceAgent Bench (2025) | 6 | Program execution rate, code similarity, task success rate, and cost | Modular tasks | 10 | Cross-disciplinary applications | ✗ | ✓ | 6 | 3 | ❌Doesn't rank module contribution (only reports "RAG improves by 14%") | ▶ Incorporates economic cost metrics |
| CellVoyager* (2025) | 1 | Result accuracy | Modular tasks | 50 | scRNA-seq analysis | ✗ | ✓ | 2 | - | ❌Doesn't analysis why CellVoyager outperformed other agents and the latent limits of current biomedical agents | ▶ Out-of-the-box creativity<br>▶ Novel biological findings<br>▶ Human-in-the-loop synergy |
| Biomni* (2025) | 1~2 for each task type | Result accuracy | QA and full workflows | 11 | Causal gene prioritization, drug repurposing, rare disease diagnosis, micro-biome analysis, and molecular cloning | ✓ | ✓ | 1 | - | ❌Doesn't analysis why Biomni outperformed other agents and the latent limits of current biomedical agents | ▶Integrate 150 tools + 105 packages + 59 databases<br>▶Generalist yet expert-level: +20-400 % vs baselines, human-par MCQ |
| SpatialAgent* (2025) | 1~2 for each task type | Result accuracy | Full workflows | 3 for spatial biology | Gene panel design, cell and niche annotation, and cell-cell interaction analysis | ✓ | ✓ | 1 | - | ❌Doesn't analysis why SpatialAgent outperformed other agents and the latent limits of current biomedical agents | ▶Outperforms strongest spatial-omics baselines<br>▶End-to-end in 30 min vs. hours of expert labor<br>▶Zero-shot generalisation |
| Ours | 18 | Cognitive program synthesis, collaboration and execution efficiency, knowledge integration and task completion quality | Full workflows | 50 | Diverse single-cell omics analysis scenarios, like spatial deconvolution and multi-omics integration | ✓ | ✓ | 8 | 3 | ✅ Atribution analysis:<br>• Prompt variation and agent robustness<br>• Functional module contribution<br>• Failed tasks | ▶Pioneering single-cell agent quantitative benchmark<br>▶ End-to-end task coverage<br>▶ Pluggable across agents/LLMs<br>▶Groundbreaking insights |

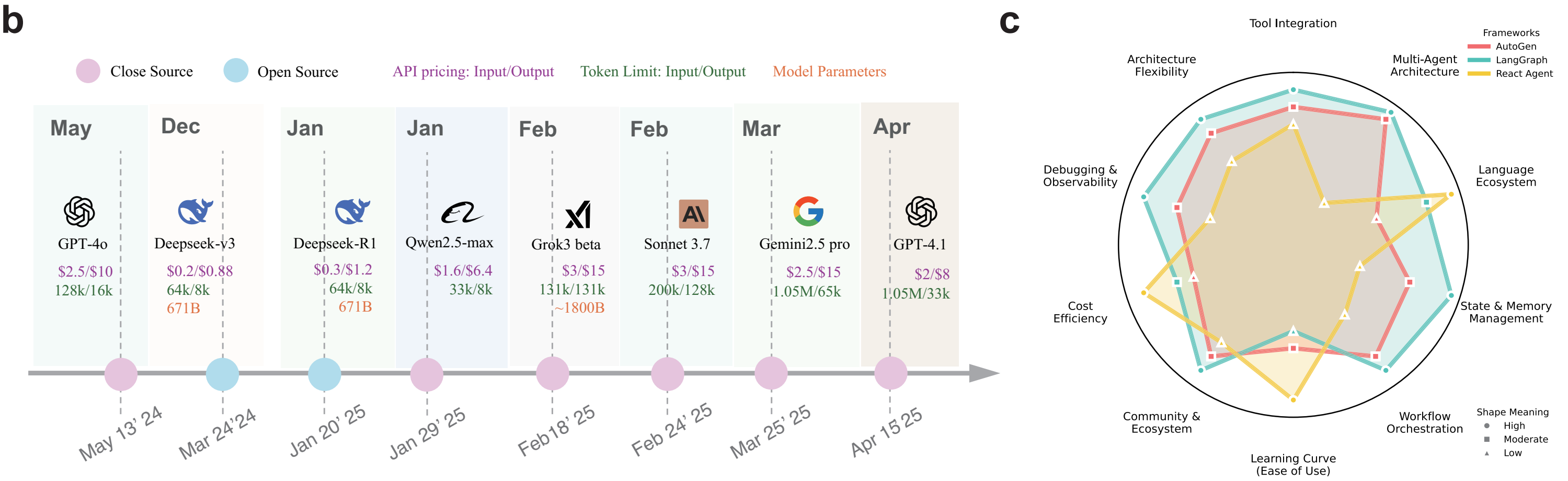

**Fig. S1. The features of current agent benchmarks in bioinformatics, LLMs and agent frameworks. a. Comparison of agent benchmarks in bioinformatics.** The figure contrasts our proposed benchmark with prior work across five key dimensions: Evaluation Metrics, Task Coverage, Platform Universality, Core Capability Insights, and Strengths. Note that # denotes *Number*. * represents works for proposing new agent architectures, rather than on traditional benchmarking studies. **b. LLMs evaluated in this study.** Eight widely used LLMs were selected in the evaluation, including GPT-4o, GPT-4.1, DeepSeek-R1, DeepSeek-V3-0324 (referred as DeepSeek-V3), Qwen-2.5-max, Sonnet-3.7, Gemini-2.5-pro, Grok3-beta. **c. Agent frameworks evaluated in this study.** The radar chart illustrates the characteristics of the three representative agent frameworks (i.e., ReAct, LangGraph, and AutoGen).

a

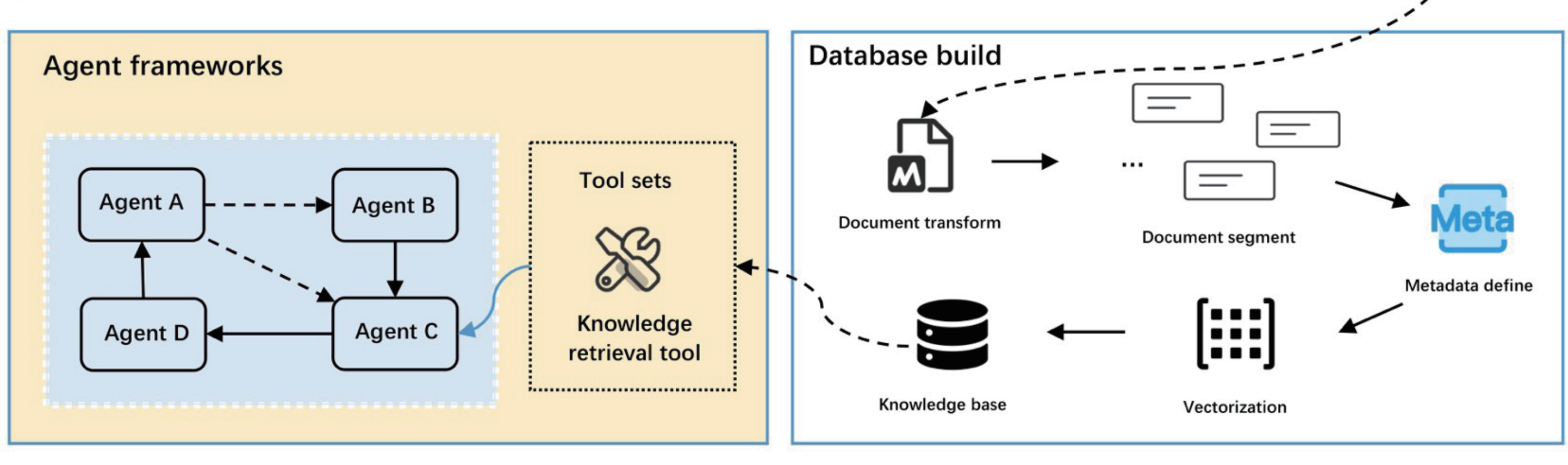

b

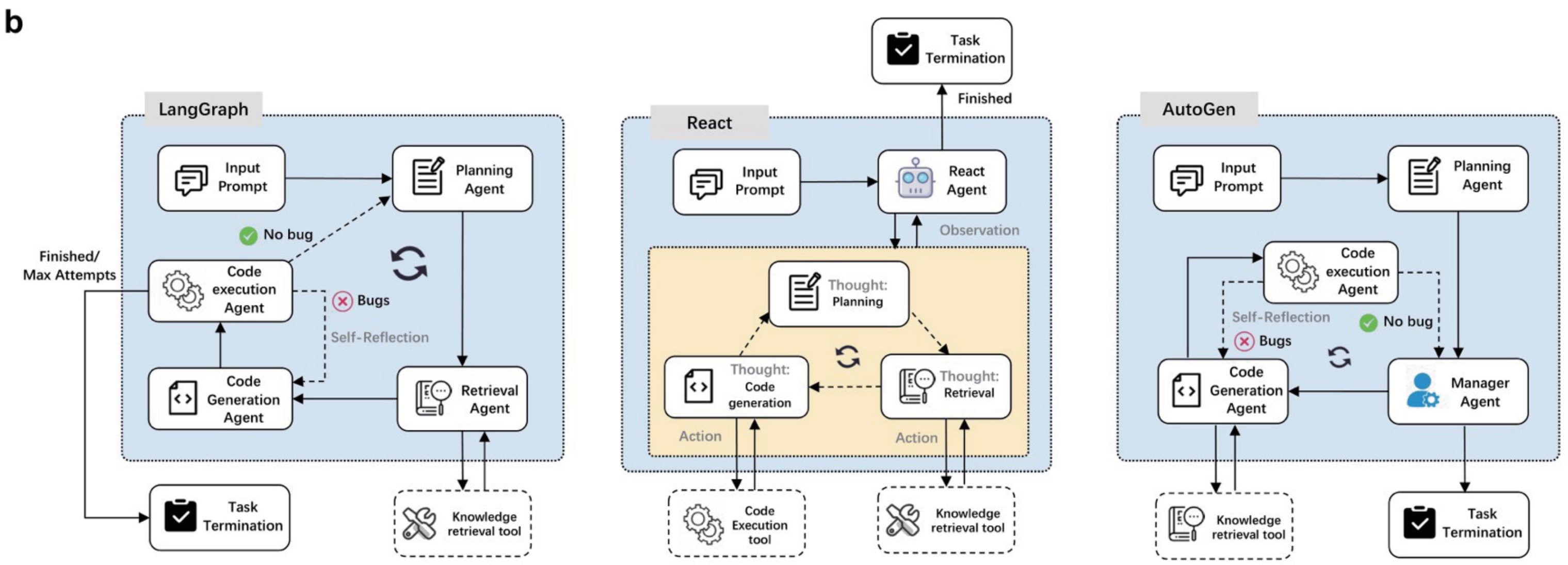

**Fig. S2. Technical details of the main experiment. a. The preprocessing of the benchmarking tasks.** Publicly available web content and scientific literature on bioinformatics tools and algorithms were collected, filtered, and converted into a semi-structured format. The resulting knowledge units were embedded using OpenAI's text-embedding-3-large model and stored in a Chroma vector database for semantic retrieval and querying. **b. Detailed architectures of multiple agent frameworks.** AutoGen consists of a Planner, Coder, Executor, Task Manager, and a retrieval-augmented bioinformatics documentation library. LangGraph comprises multiple agents-planner, code generation, execution, and retrieval agents coordinated in a structured flow with a capped number of iterations. ReAct adopts a single-agent design following a "thought-action-observation" loop. The agent integrates planning, tool use, and result evaluation, and performs retrieval, code generation, and execution in flexible reflection cycles.

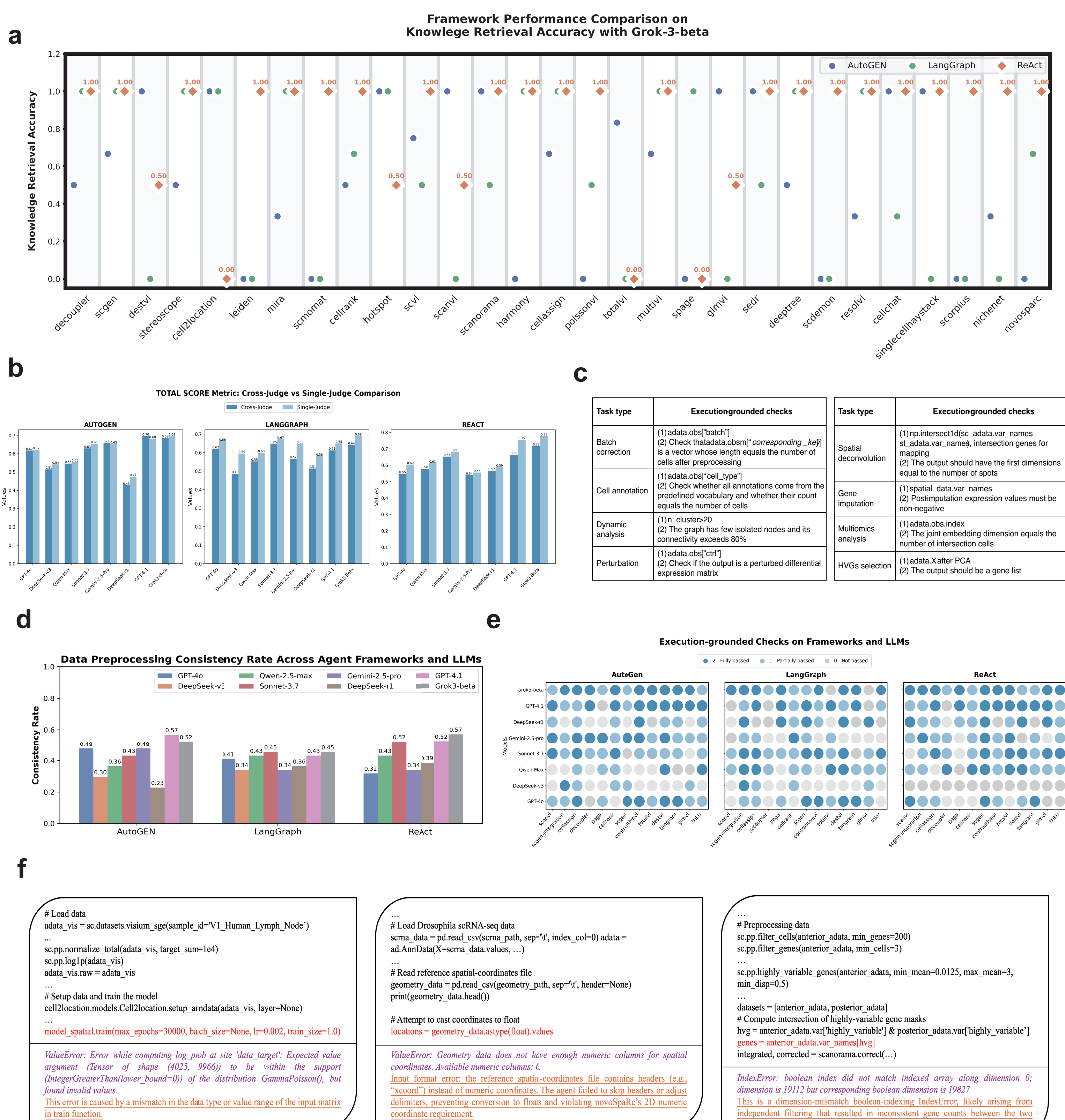

**Fig. S3. a. Framework performance comparison on knowledge retrieval accuracy with Grok3-beta.** The figure shows knowledge retrieval accuracy for the three agent frameworks on detailed tasks where retrieval results differed (tasks with identical results were excluded). **b. Comparison of single-judge scoring versus cross-judge scoring by LLMs in total scores. c. Execution-grounded checks for eight representative task types.** Two checks are selected manually for each task type. **d. Unit testing for the data preprocessing stage.** If the preprocessed data from the agent's run and the groundtruth script have the same number of cells*genes, the unit result is considered consistent; otherwise, it is considered inconsistent. **e. Execution-grounded checks across agent frameworks and LLMs.** Dark green indicates both checks passed, light green indicates only one check met the condition, and gray indicates neither check passed. **f. Key code errors encountered by Grok3-beta under the AutoGEN framework.** Analysis of Cell2location, Scanorama and NovoSpaRc failures (from upper to down) reveals that critical errors predominantly originate during data processing.

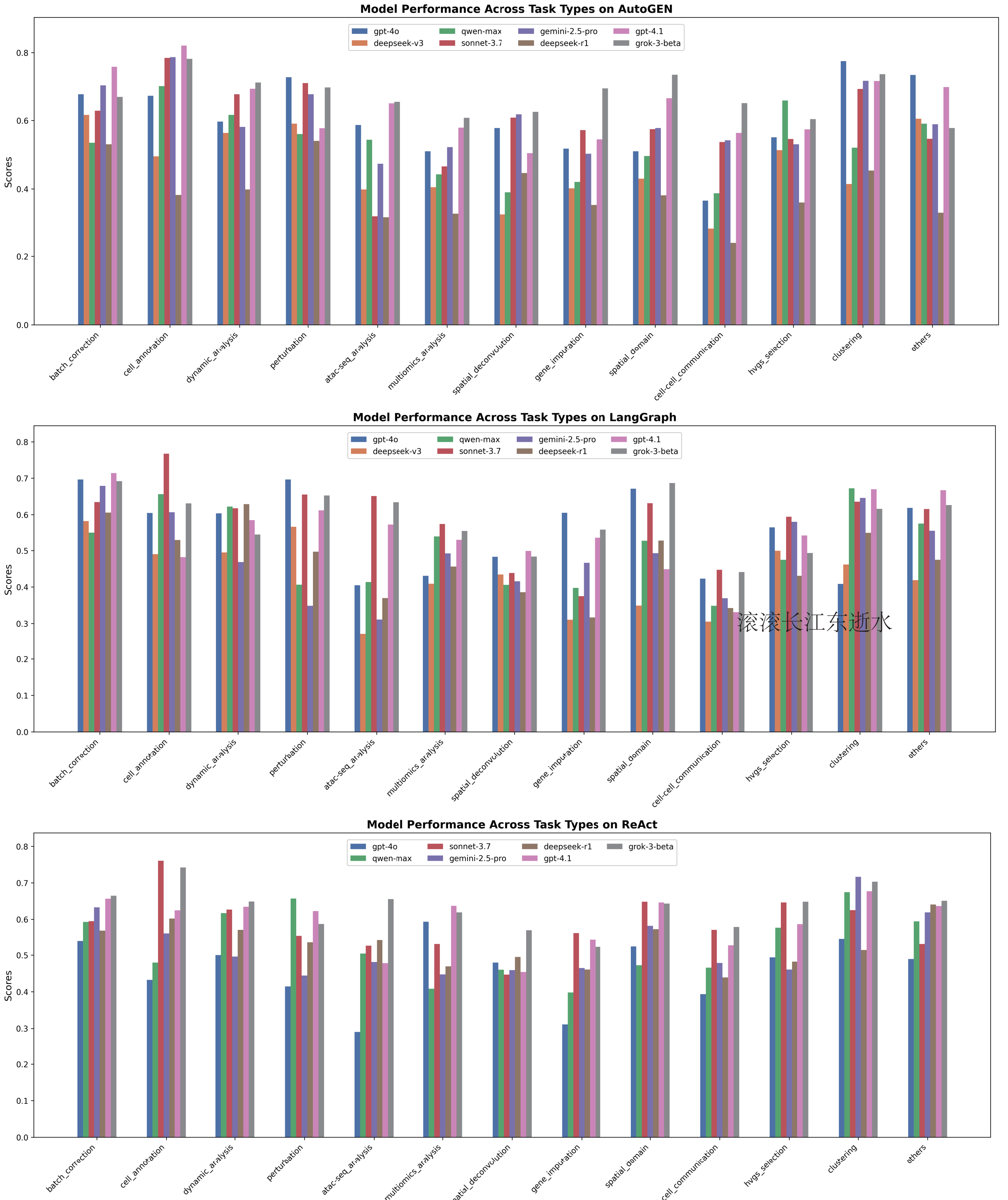

**Fig. S4. Model performance across task types on three agent frameworks.** The bar plot displays the total scores of different LLMs across the three agent frameworks—AutoGEN, LangGraph, and ReAct—by task type. These total scores are derived from 17 evaluation metrics in the study.

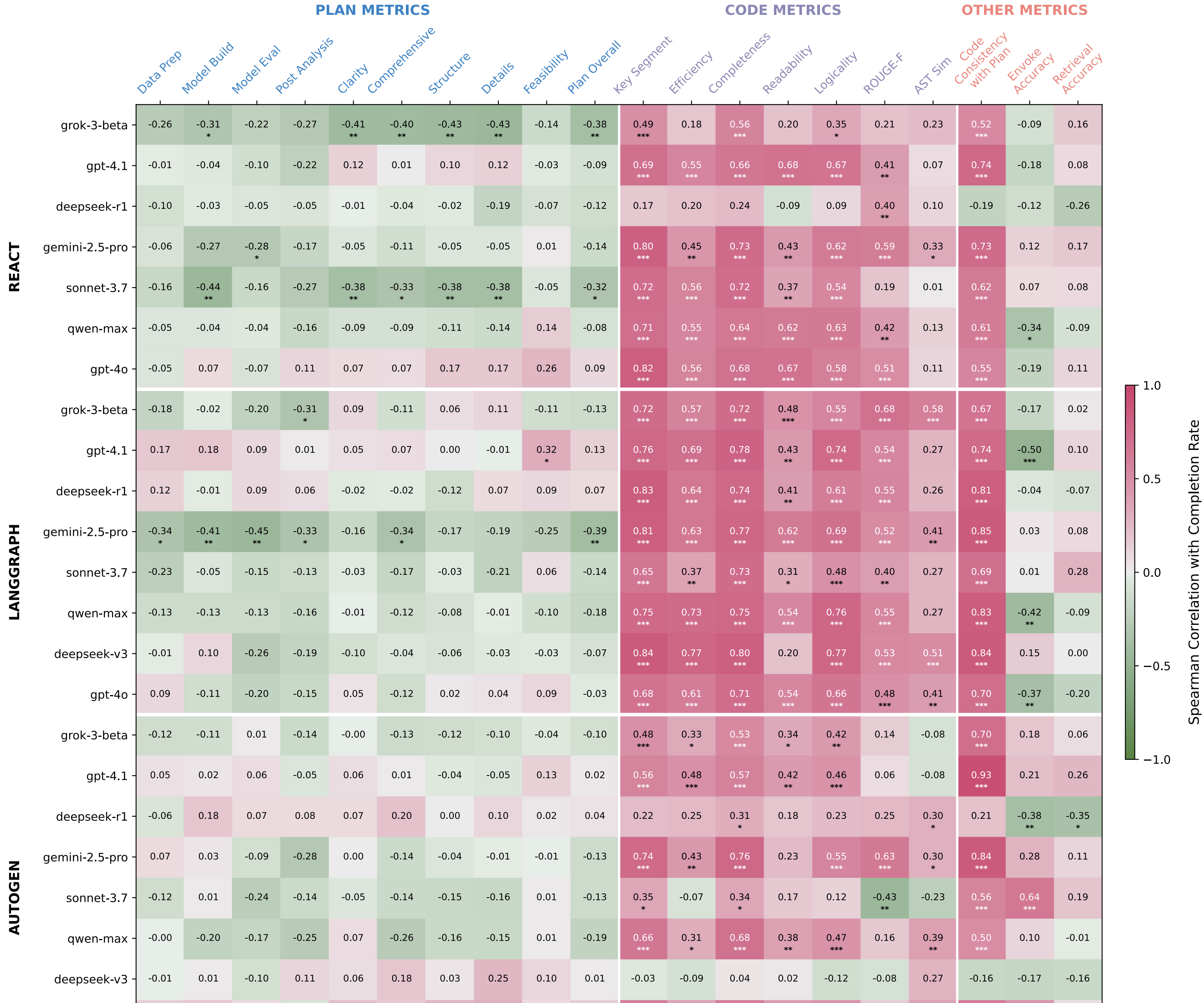

**Fig. S5. Spearman correlation among agent performance metrics for main experiment.** Heatmap shows the Spearman correlation structure among task completion rate, plan metrics, code metrics, and other metrics, aggregated across evaluations of three agent frameworks and eight LLMs. Task completion rate is positively correlated with code quality metrics and code consistency with plan across most models and frameworks.

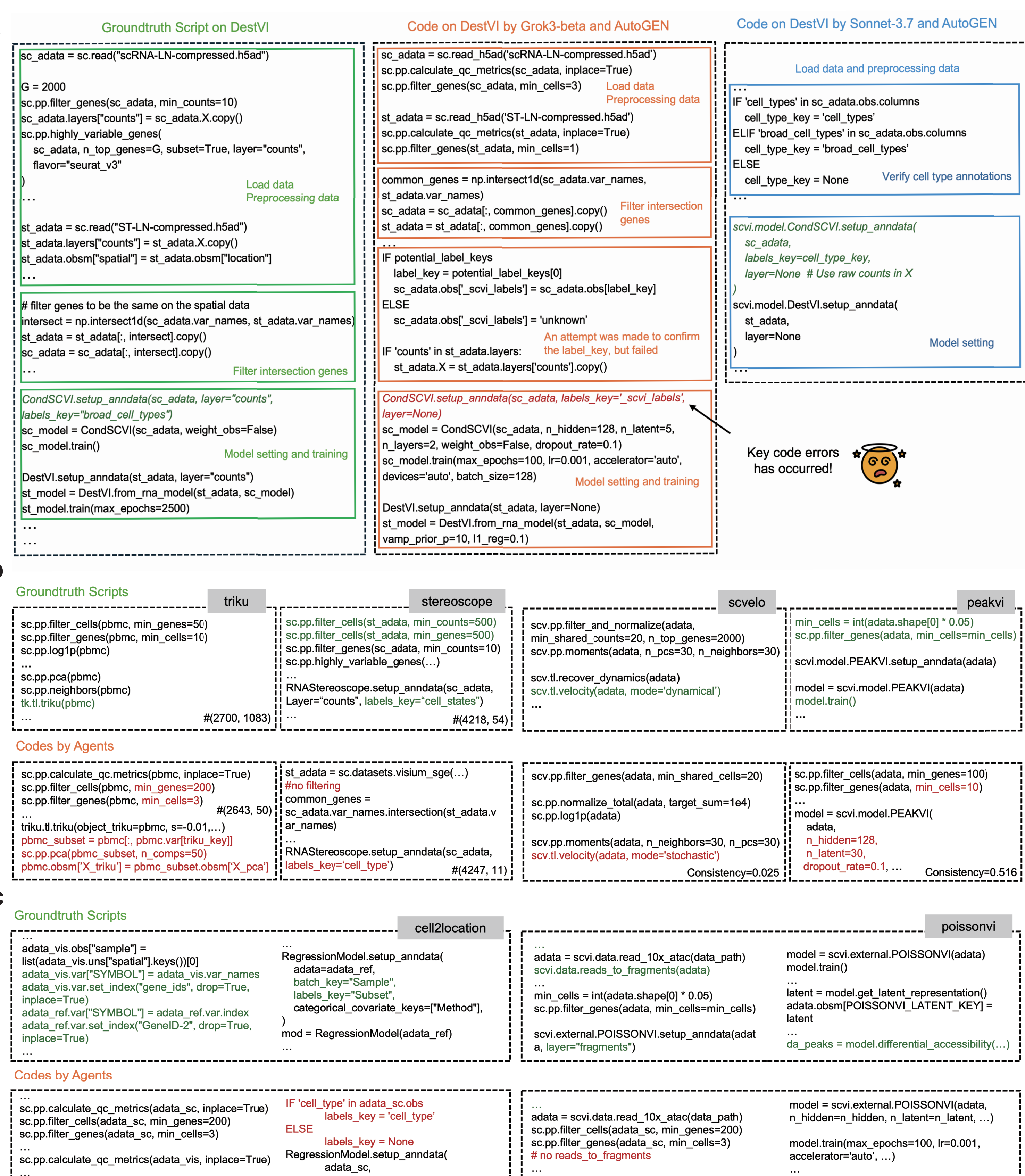

**Fig. S6. Comparisons between the code generated by agents and the ground-truth code. Green indicates correct code snippets, while red indicates incorrect code snippets generated by the agent. a. A detailed example of code comparison on DestVI.** Compare the ground truth code with the code generated by the agent to locate the key snippet responsible for the task failure. **b. Cases on shape mismatch and low consistency results.** Four code snippets from the triku, stereoscope, scvelo, and peakvi tasks, each highlighting a critical comparison. **c. Cases on failure tasks.** Two code snippets from the cell2location and poissonvi tasks, each highlighting a critial comparison.

**a**

## Batch correction (Scanorama)

Tutorial (Groundtruth)

AutoGEN+Gemini-2.5-pro

LangGraph+Gemini-2.5-pro

ReAct+Grok3-beta

clusters

library_id

V1_Mouse_Brain_Sagittal_Anterior

V1_Mouse_Brain_Sagittal_Posterior

## Spatial deconvolution (Tangram)

Tutorial (Groundtruth) — EC5_art

AutoGEN+Grok3-beta — Endothelial

LangGraph+GPT-4.1 — Endothelial

ReAct+GPT-4.1 — Endothelial

spatial1

spatial2

**b**

Correlation of Two Metrics on CellAssign

Pearson r = 0.653 (p = 0.1116)
Spearman ρ = 0.549 (p = 0.2014)

Confusion Matrix

Result Consistency

Correlation of Two Metrics on scGEN

Pearson r = 0.647 (p = 0.0068)
Spearman ρ = 0.740 (p = 0.0010)

Pearson

Result Consistency

Correlation of Two Metrics on SpaGE

Pearson r = 0.953 (p = 0.0475)
Spearman ρ = 0.949 (p = 0.0513)

Spearman

Result Consistency

Correlation of Two Metrics on Triku

Pearson r = 0.495 (p = 0.0858)
Spearman ρ = 0.318 (p = 0.2895)

Jaccard@2k

Result Consistency

**c**

| **Tasks** | Batch correction | Clustering | Spatial deconvolution | HVGs selection | Gene imputation | Cell Annotation | Trajectory inference | Pertubation |
|---|---|---|---|---|---|---|---|---|
| **Biological Metric** | Aggregated score | ARI/NMI | Pearson correlation | Jaccard@2k | Spearson correlation | Confusion matrix | Graph connectivity | Pearson correlation |

**Fig. S7. a. Visualization of biological results.** Batch correction (Scanorama) and Spatial deconvolution (Tangram) were selected to show the results produced by the groundtruth script (from tutorial) and the results produced by the agents. **b. Correlations between result consistency and the biological metric on CellAssign, scGEN, SpaGE and Triku.** CellAssign, scGEN, SpaGE and Triku are tools for cell annotation, pertubation prediction, gene imputation and HVGs selection, respectively. Their corresponding biological metrics are confusion matrix, pearson correlation, spearman correlation and Jaccard@2k, respectively. **c. Specific tasks and the corresponding biological metric.**

| | Basic | Intermediate | Advanced |
| --- | --- | --- | --- |
| **Brief description** | Use CellTypist for automated cell type annotation of single-cell RNA sequencing data. | The same as basic | The same as basic |
| **Dataset location** | Dataset location:<br>- Input data for Training: '/path/to/demo_2000_cells.h5ad'<br>- Input data for Testing: '/path/to/demo_400_cells.h5ad' | The same as basic | The same as basic |
| **Result saving** | Must-save Results:<br>- Make sure cell type annotation result of celltypist is stored in adata.obs['predicted_labels'] and save adata.obs['predicted_labels'] as 'celltypist.csv' | The same as basic | The same as basic |
| **Core analysis steps** | - | - | Core Analysis Steps:<br>1.Load the single-cell data using Scanpy<br>2.Perform cell type annotation using CellTypist with majority voting<br>3.Convert predictions to AnnData format |
| **Key requirements** | - | Key Requirements:<br>- Use CellTypist annotation method<br>- Apply majority voting for predictions | The same as intermediate |
| **Libraries** | - | - | Libraries to Use:<br>- Scanpy<br>- celltypist |

**Fig. S8. Prompt templates.** Three prompt templates--basic, intermediate, and advanced--were designed to assess the robustness of agent performance under varying input conditions, using celltypist task as an example. The intermediate and advanced templates extend the basic version by incorporating key requirements, core analysis steps, and required li-braries. These templates allow for systematic evaluation of input prompt effects across different agent frameworks.

**a**

Tangram w/ basic prompt
Load spatial and scRNA-seq data
Preprocess the data
Run Tangram mapping
Save the results
Verify the output

Tangram w/ intermediate prompt
Load spatial and scRNA-seq data
Preprocess the data
Create the MuData object
Error retrieval!
Identify overlapping genes between spatial and scRNA-seq
Error retrieval!
Define the image layer for segmentation
Calculate segmentation features
Run Tangram mapping
Save the results
Verify the output
An AttributeError occurred on squidpy's calculate_image _features function
ValueError: Please set up your AnnData with Tangram.setup_anndata
'rna_count_based not found in adata.obs.
...
Fail

Decoupler w/ basic prompt
Load scRNA-seq
Load marker genes
Error retrieval!
Preprosess the data
Use Decoupler to annotate cells
Save Results
Save Results

Decoupler w/ advanced prompt
Load and preprocess scRNA-seq
Dimensionality reduction and clustering
Load marker genes
Error retrieval!
Pydeseq2
Run Decoupler's ORA Method for enrichment analysis
Process ORA Results to Handle Infinity Values
Rank cell types per cluster with t-test method
Assign Cell Type Annotations to Clusters
Save the results
KeyError: 'pct_counts_mt'
TypeError: get_ora_df() got an unexpected keyword argument 'mat'
'dict' object has no attribute 'columns'
...
Fail

**b**

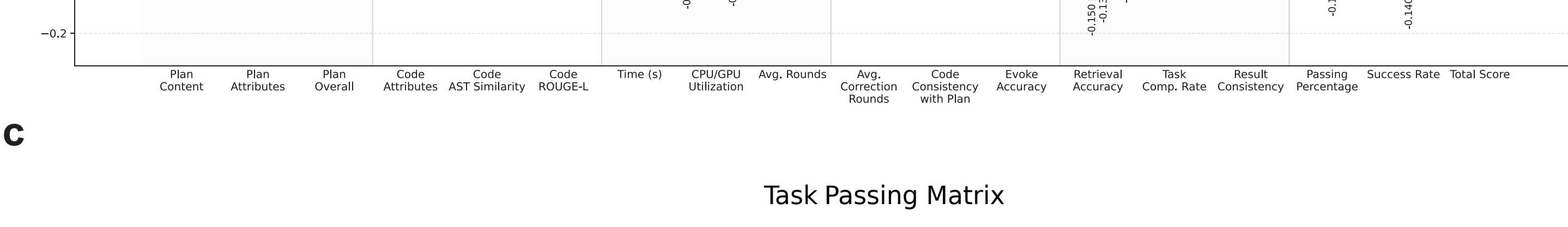

**c**

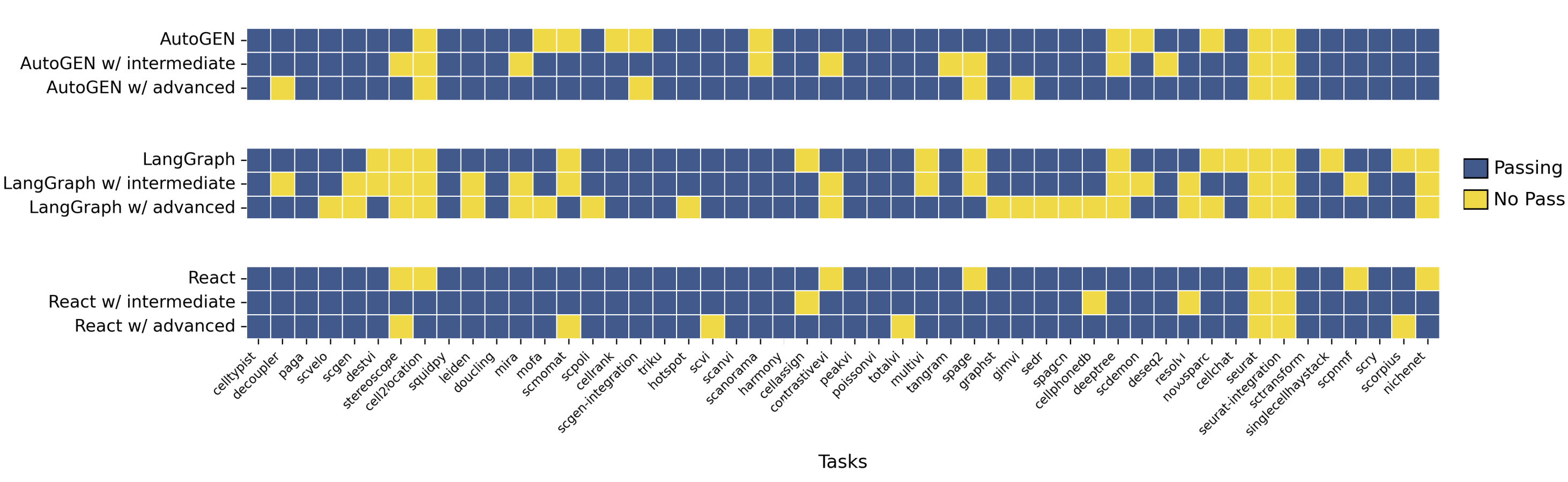

**Fig. S9. Detailed information for Prompting Strategies. a. Workflow comparison for intermediate/ advanced prompt failures.** Left: AutoGEN (on Grok-3 Beta) executing the Tangram task: Basic Prompt vs. Intermediate Prompt.
Right: AutoGEN (on Grok-3 Beta) executing the Decoupler task: Basic Prompt vs. Advanced Prompt. **b. Comparison of metrics differences with intermediate/advanced prompting strategies on three agent frameworks.** Metrics are first mapped into [0, 1] with formulae presented in Supplementary Method. For Time, GPU/CPU Utilization, Avg. Rounds and Avg. Correction Rounds, higher values indicate less time consumed, lower resource usage, and fewer rounds required. **c. Failure heatmap.** A visual comparison of task pass rates across the original AutoGEN, LangGraph, and ReAct frameworks alongside their performance with prompt tiers.

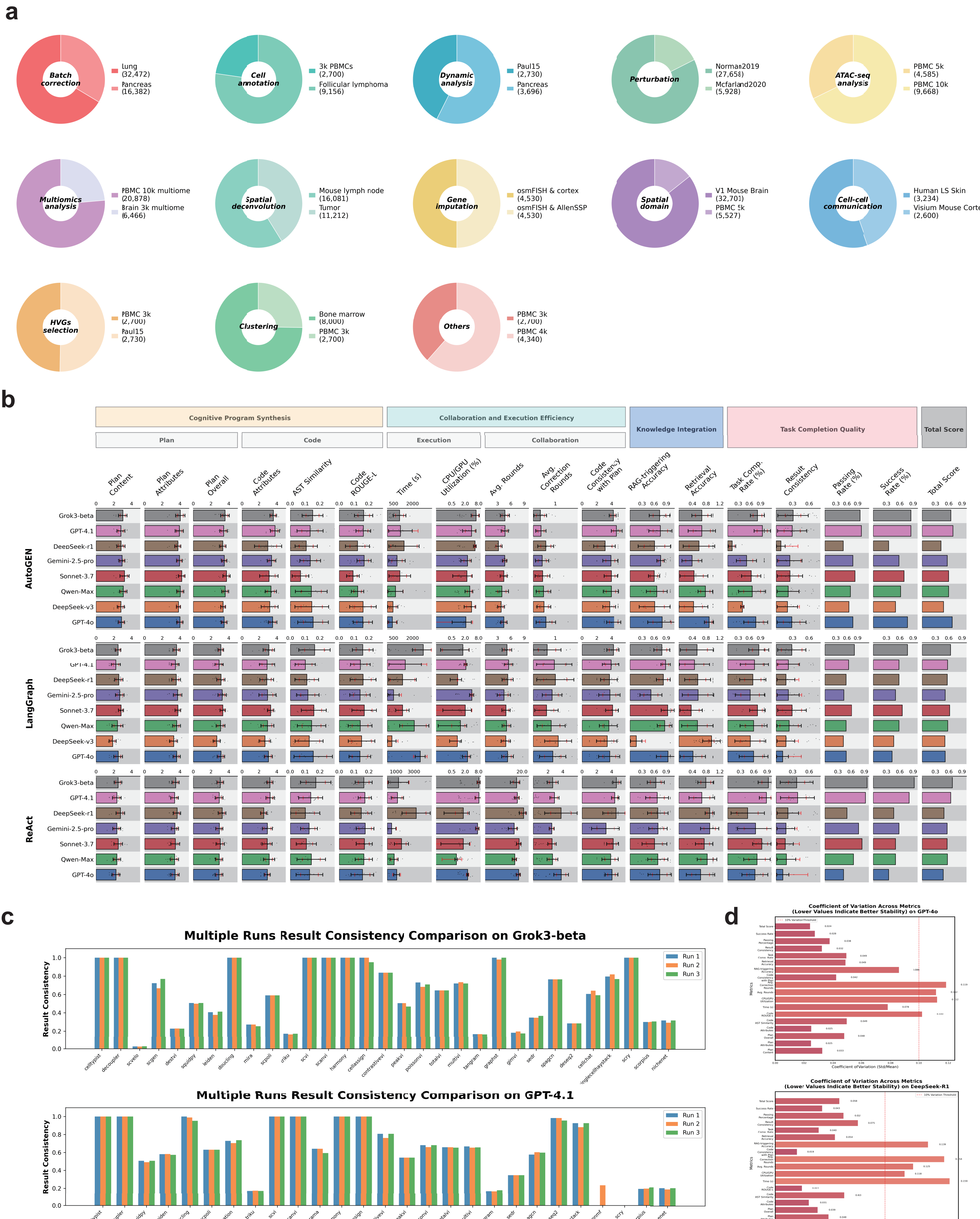

**Fig. S10. a. Dataset information for datasets variation analysis.** An additional dataset for one subtask/tool is collected under each major task category (13 categories in total). Dataset names and cell/spot count of two distinct sets of datasets for each task type are shown in the figure. **b. Results on the multiple datasets across agent frameworks and LLMs.** All metrics in the benchmark are computed and shown as bar plots. Individual measurements from 26 datasets per framework-LLM combination are depicted as scatter points, error bars represent ±1 standard deviation (black) and 95% CI (red) of these measurements. **c. Result consistency of multiple runs on Grok3-beta and GPT-4.1. d. Variation across metrics of multiple runs on different LLMs.** The red dashed line in the figure represents the 10% variation boundary.

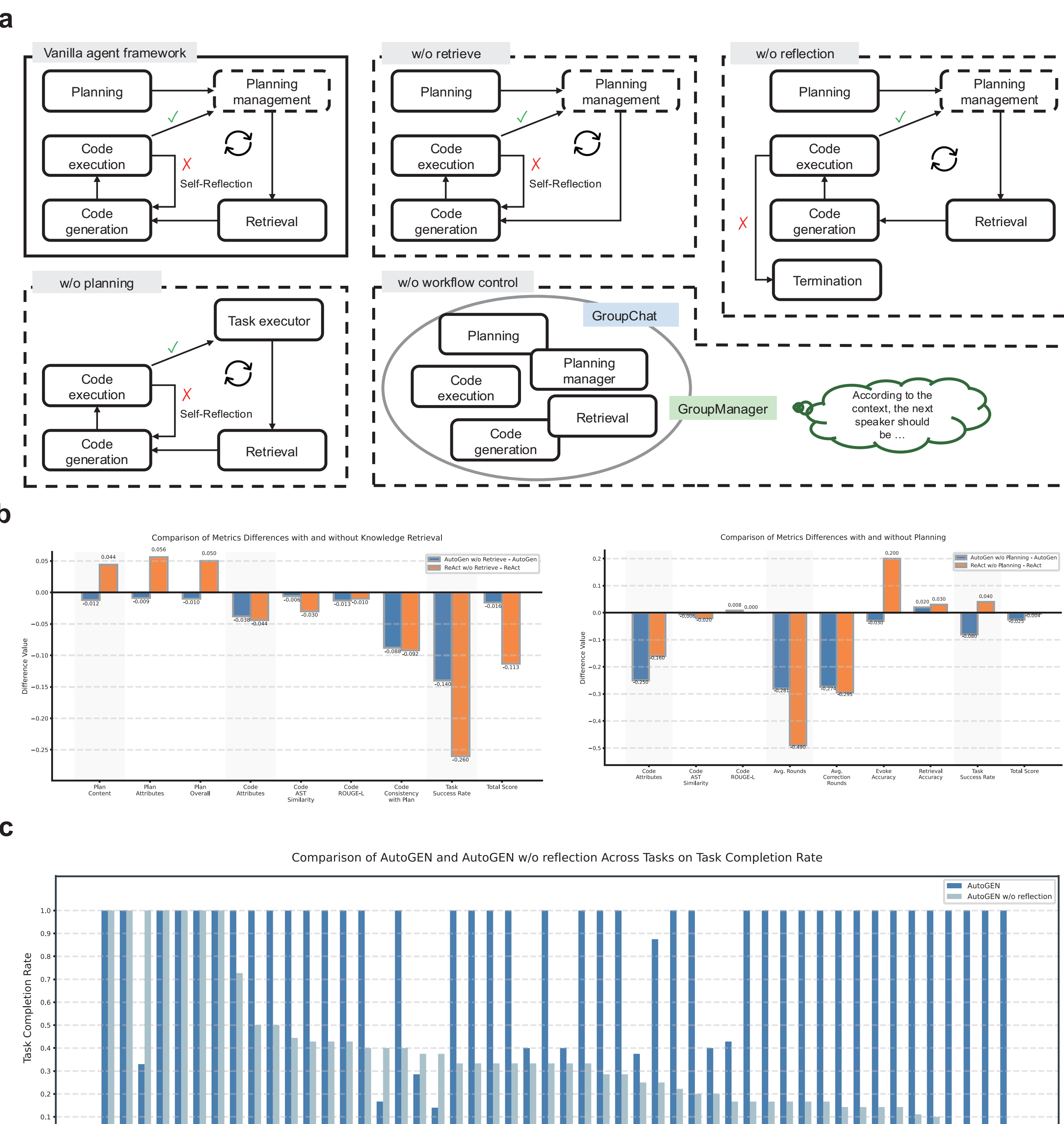

**Fig. S11. a. Illustration of different functional module ablations in agent frameworks.** The vanilla agent framework denotes the structure employed in the main experiments. w/o retrieve indicates the absence of the retrieval capability. w/o reflection indicates the absence of the self-reflection module. w/o planning indicates the absence of the pre-execution planning module. In AutoGen, this was implemented by replacing planner agent with a task executor agent, while in ReAct, planning was disabled by eliminating all planning-related instructions from the system prompt. w/o inter-agent workflow control indicates the absence of workflow control where the group manager autonomously determines the next speaker. **b. Detailed information for ablation analysis.** Comparison of metrics differences on knowledge retrieval and planning ablation with AutoGEN and ReAct. Metrics are first mapped into [0, 1] with formulae presented in Supplementary Method. For Avg. Rounds and Avg. Correction Rounds, higher values indicate fewer rounds required. **c. Task completion rate on reflection ablation across 50 tasks.** Tasks are ordered by the performance of AutoGEN w/o reflection from highest to lowest.

**b**

| Error Types | Data Prep | Model Build | Model Eval | Post Analysis | Clarity | Comprehensive | Structure | Details | Feasibility | Key Segment | Efficiency | Completeness | Readability | Logicality | ROUGE-F | AST Sim | Code Consistency with Plan | Task Comp. Rate | Envoke Accuracy | Retrieval Accuracy |
|---|---|---|---|---|---|---|---|---|---|---|---|---|---|---|---|---|---|---|---|---|
| | PLAN METRICS | | | | | | | | | CODE METRICS | | | | | | | OTHER METRICS | | | |
| Inconsistent Planning | 0.02 | 0.06 | 0.06 | 0.09** | 0.01 | 0.06 | 0.01 | 0.01 | -0.04 | -0.66*** | -0.47*** | -0.67*** | -0.40*** | -0.55*** | -0.39*** | -0.21*** | -0.59*** | -0.71*** | -0.03 | -0.05 |
| Planner Issue | 0.01 | 0.06 | 0.05 | 0.08** | -0.01 | 0.05 | -0.01 | -0.00 | -0.05 | -0.64*** | -0.46*** | -0.65*** | -0.40*** | -0.55*** | -0.37*** | -0.20*** | -0.57*** | -0.69*** | -0.01 | -0.05 |
| Poor Instructions | 0.02 | 0.05 | 0.05 | 0.09** | -0.00 | 0.05 | 0.01 | -0.00 | -0.06* | -0.61*** | -0.43*** | -0.62*** | -0.38*** | -0.52*** | -0.34*** | -0.18*** | -0.52*** | -0.66*** | -0.02 | -0.05 |
| Ambiguous Input | 0.05 | 0.08** | 0.09** | 0.12*** | 0.03 | 0.10*** | 0.05 | 0.05 | -0.05 | -0.54*** | -0.39*** | -0.56*** | -0.34*** | -0.47*** | -0.30*** | -0.18*** | -0.46*** | -0.60*** | -0.02 | -0.08* |
| Knowledge Failure | 0.07* | 0.07* | 0.11*** | 0.12*** | 0.10*** | 0.10*** | 0.11*** | 0.07* | 0.05 | -0.56*** | -0.39*** | -0.58*** | -0.28*** | -0.44*** | -0.37*** | -0.21*** | -0.57*** | -0.66*** | -0.04 | -0.13*** |
| Context Handling | 0.10** | 0.13*** | 0.13*** | 0.14*** | 0.03 | 0.12*** | 0.06 | 0.07* | 0.01 | -0.35*** | -0.25*** | -0.34*** | -0.27*** | -0.31*** | -0.20*** | -0.12*** | -0.19*** | -0.42*** | -0.04 | -0.06 |
| Redundant Process | 0.01 | -0.03 | 0.01 | -0.02 | 0.03 | -0.02 | 0.01 | -0.02 | 0.05 | -0.31*** | -0.25*** | -0.32*** | -0.17*** | -0.23*** | -0.21*** | -0.11*** | -0.38*** | -0.34*** | 0.03 | 0.01 |
| Coder Issue | 0.00 | -0.03 | 0.01 | -0.03 | 0.03 | -0.02 | 0.01 | -0.02 | 0.06 | -0.27*** | -0.20*** | -0.29*** | -0.12*** | -0.21*** | -0.20*** | -0.12*** | -0.34*** | -0.29*** | 0.01 | 0.01 |
| Missing Dependency | 0.00 | -0.07* | 0.04 | -0.05 | 0.05 | -0.04 | 0.04 | -0.01 | 0.02 | -0.21*** | -0.13*** | -0.21*** | -0.09** | -0.13*** | -0.11*** | -0.04 | -0.24*** | -0.22*** | -0.02 | -0.06* |
| Lack Clarification | 0.03 | 0.02 | 0.03 | 0.05 | 0.05 | 0.06* | 0.04 | 0.06* | 0.05 | -0.19*** | -0.18*** | -0.21*** | -0.13*** | -0.17*** | -0.18*** | -0.13*** | -0.24*** | -0.22*** | 0.05 | 0.04 |
| Early Stopping | 0.04 | 0.01 | 0.07* | 0.03 | 0.02 | 0.03 | 0.01 | 0.03 | -0.00 | -0.11*** | -0.04 | -0.10** | -0.05 | -0.04 | -0.02 | -0.02 | -0.16*** | -0.13*** | 0.01 | -0.01 |
| Time Limit | 0.04 | 0.03 | 0.02 | 0.04 | -0.01 | 0.00 | -0.01 | 0.00 | -0.02 | -0.02 | -0.04 | -0.04 | -0.07* | -0.05 | 0.01 | -0.01 | -0.00 | -0.03 | -0.00 | 0.01 |
| Interrupted Exec | -0.04 | -0.06* | -0.03 | -0.05 | -0.05 | -0.07* | -0.08** | -0.08* | -0.08** | -0.04 | -0.04 | -0.04 | -0.05 | -0.04 | -0.06 | -0.02 | -0.01 | -0.04 | 0.01 | -0.01 |
| Info Not Shared | -0.02 | 0.01 | -0.00 | 0.00 | 0.02 | 0.01 | 0.01 | 0.01 | -0.02 | 0.01 | -0.05 | 0.02 | -0.03 | -0.00 | -0.05 | 0.03 | 0.03 | -0.02 | -0.01 | 0.02 |

**Fig. S12. a. Cumulative distribution of error types across tasks.** Heatmap shows the cumulative frequency of 14 predefined error types across 50 single-cell omics analysis tasks. The distribution patterns indicate variation in the prevalence of specific error types across different task categories. **b. Spearman correlation between error types and evaluation metrics.** Spearman correlation coefficients were computed between 14 error types and a range of performance metrics, including plan metrics, code metrics, and other metrics. The heatmap displays correlation values for all error types, with the top 10 most frequent errors showing negative correlations with task completion rate, code metrics, and code consistency with plan.

| Parameters | Values | Definition |
| --- | --- | --- |
| Execution Time: $s_t = 1 - \frac{log_b(t+c)}{log_b(T+c)}$ | | |
| $t$ | - | real time by agents |
| $c$ | 100 | offset |
| $b$ | 1.8 | logarithm base |
| $T$ | 10800 | a constant, maximum time |
| CPU/GPU utilization: $s_{ru} = \frac{e^{-i\cdot s_{ru}^{j}} - e^{-i\cdot max^{j}}}{1-e^{-i\cdot max^{j}}}$ | | |
| $s_{ru}$ | - | real utilization by agents |
| $i$ | 0.03 | attenuation coefficient |
| $j$ | 1.5 | declining value area enhancement index |
| $max$ | 10 | a constant, maximum utilization |
| $e$ | $e$ | a constant |
| Rounds: $s_{tr/scr} = \frac{e^{-k\cdot s_{tr/scr}^{m}} - e^{-k\cdot max^{m}}}{1-e^{-k\cdot max^{m}}}$ | | |
| $s_{tr/scr}$ | - | real rounds by agents |
| $k$ | 0.003 | attenuation coefficient |
| $m$ | 1.5 | declining value area enhancement index |
| $max$ | 50/10 | a constant, maximum rounds |
| $e$ | $e$ | a constant |
| Total Score | | |
| $w_1$ | 0.15 | the weight for cognitive program synthesis |
| $w_2$ | 0.15 | the weight for collaboration and execution efficiency |
| $w_3$ | 0.2 | the weight for bioinformatics knowledge integration |
| $w_4$ | 0.5 | the weight for task completion quality |

**Fig. S13.** Parameters in equations of execution time, CPU/GPU utilization, rounds and total score.

## Text S1. Comparisons of single-judge scoring and cross-judge scoring with LLMs

To mitigate potential biases in evaluation metrics (such as Plan Content and Plan Attribute) when using LLMs for scoring, we employed three distinct LLMs—GPT-4o, Grok3-beta, and Gemini-2.5-pro—and used the average of their scores as the final result. Besides, we observed that LLM-based scoring exhibits larger bias in evaluating RAG-triggering accuracy, with a tendency for models to assign higher scores to their own outputs. To address this, we selected 13 task subsets and incorporated expert human evaluation for three scoring LLMs. The final score of the scoring LLMs was computed as a weighted combination: 0.8 × LLM score + 0.2 × expert score, thereby reducing model-induced bias. The agreement rate between human experts and cross-judge LLM scoring on the selected subset of tasks was approximately 94%. Furthermore, on the full task set, the inter-judge agreement rates among LLMs for the LLM-graded metrics—Plan Content, Plan Attributes, Code Attributes, Code Consistency with Plan, and RAG-triggering accuracy—were 93%, 92%, 95%, 98%, and 90%, respectively.

Notably, we find that compared to single-judge scoring, the use of multi-judge scoring leaves the main conclusions of the paper unchanged. For example, the results continue to demonstrate GPT-4.1 and Grok3-beta as top performers for AutoGEN, Sonnet-3.7 and Grok3-beta for LangGraph, and Grok3-beta and GPT-4.1 for ReACT, with Grok3-beta consistently achieving optimal task success rates across all frameworks. Conclusions unrelated to LLM-based evaluation metrics—such as interaction rounds and retrieval accuracy—also remain consistent with the original findings. The key changes observed are as follows: (1) For individual evaluation metrics, multi-judge scores are generally lower than single-judge scores. This is attributed to the fact that another two judge LLMs, i.e., Gemini-2.5-pro and Grok3-beta, usually score lower scores than GPT-4o when assessing plans, codes or other parts. (2) Regarding RAG-triggering accuracy, single-judge scoring resulted in inflated performance for GPT-4o, which previously achieved the highest values across all three agent frameworks. After switching to cross-judge scoring, Grok3-beta, GPT-4.1, and DeepSeek-r1 emerge as the top-performing models in their respective agent frameworks.

To provide the reader with a clear comparison, we visualize the specific values of single-judge vs. multi-judge averaging across corresponding evaluation metrics for the three agent frameworks (i.e., AutoGEN, LangGraph, and ReAct) using bar charts as follows:

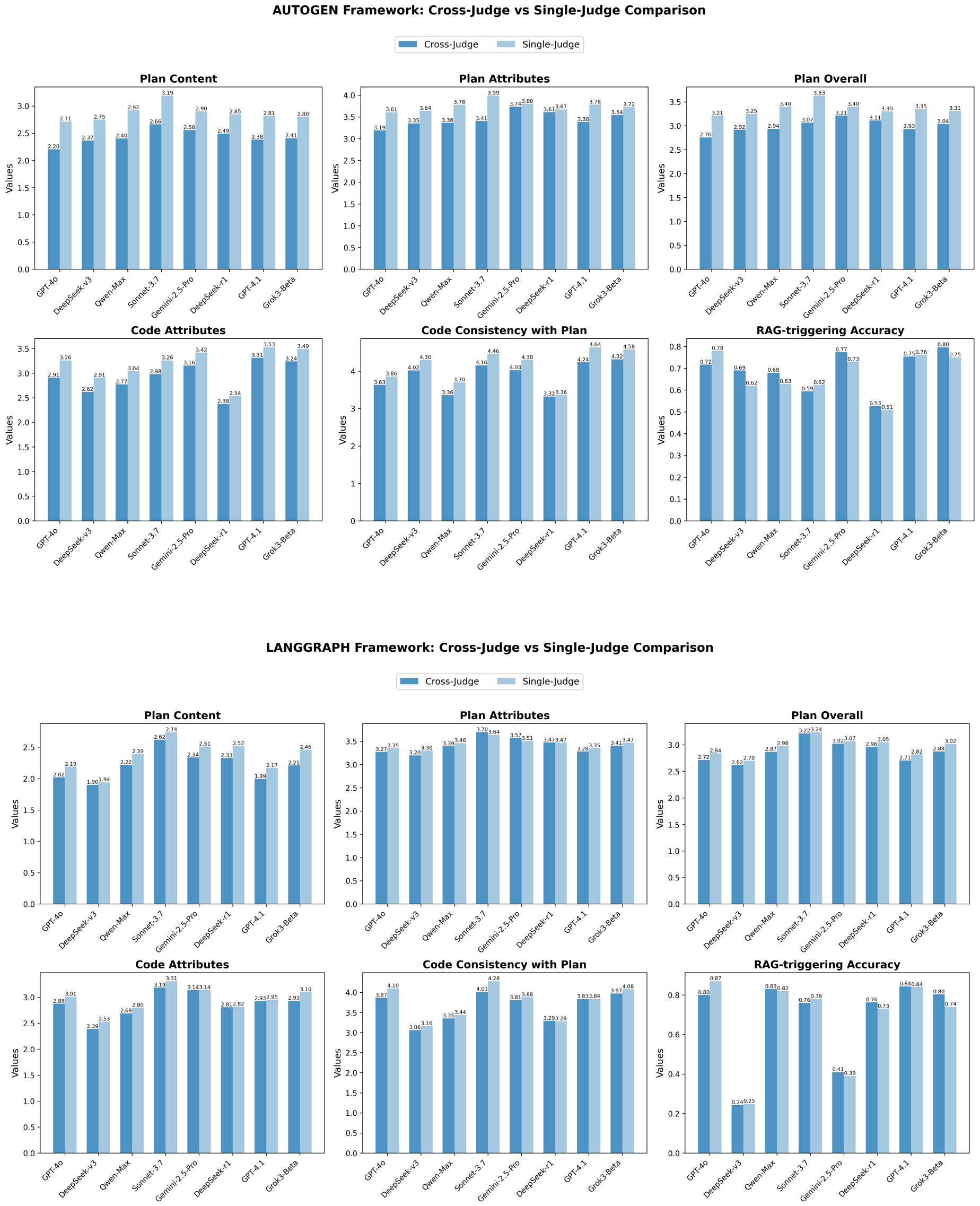

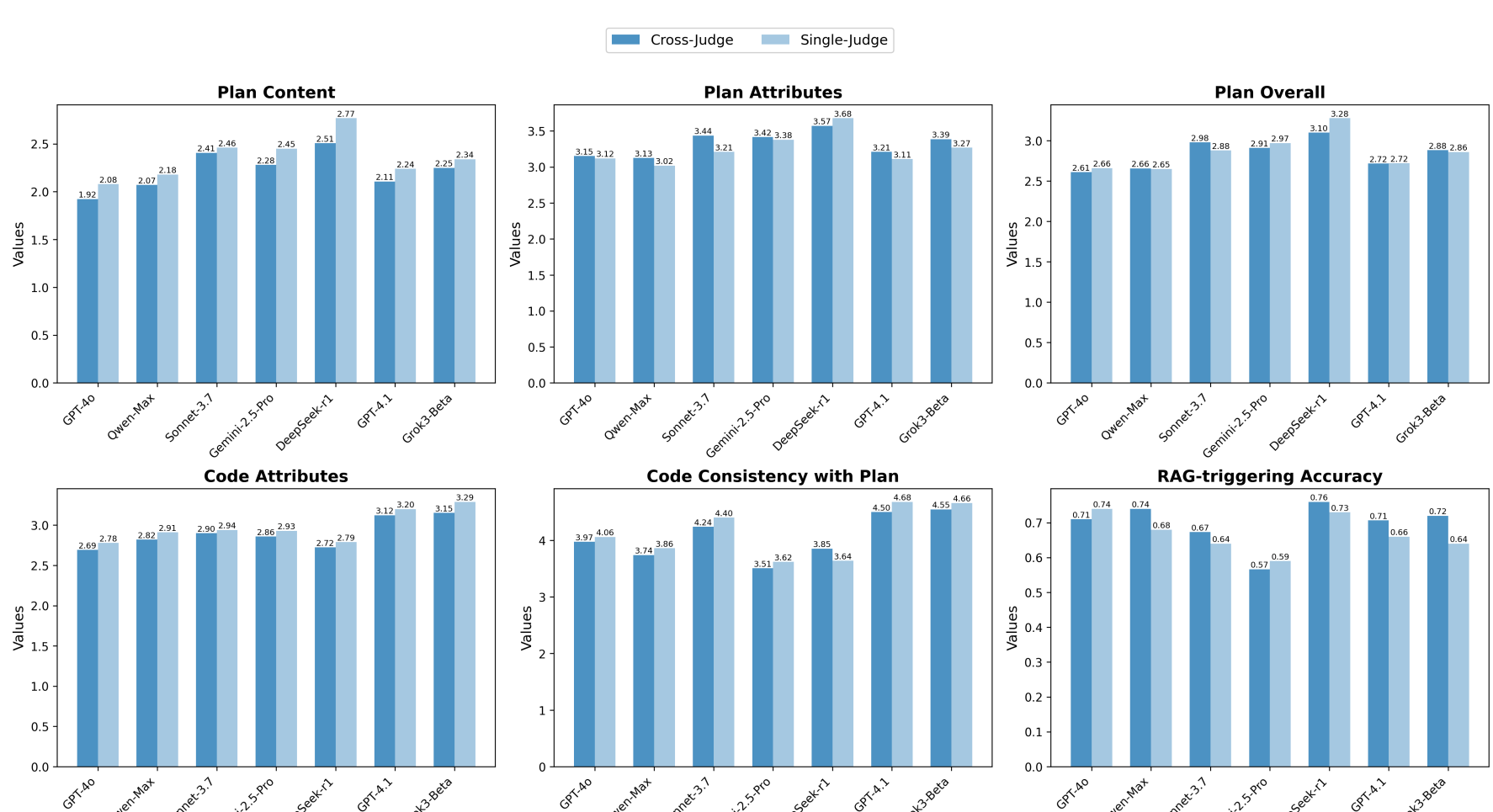

Additionally, we present the total score comparisons between cross-judge and single-judge evaluations in Additonal file 1: Fig. S3b. Regarding the total score, the ranking of agent framework + LLM combinations remains unchanged for AutoGEN, while slight changes are observed in LangGraph and ReAct (e.g., in LangGraph, DeepSeek-V3 shifts from 7th to 8th, and in ReAct, GPT-4o moves from 5th to 6th). Importantly, these adjustments are minor and do not result in LLMs previously ranked near the bottom rising to the top. This stability likely stems from the relatively low weight assigned to LLM-dependent evaluation metrics in the total score. Consequently, the variations observed in these LLM-judged metrics are further mitigated when computing the total score. As a result, the core conclusions of our study are not affected (such as top-2 LLM performers for each agent framework) by the shift to cross-judge scoring.

Furthermore, since our robustness checks (tied prompt, multi-dataset, multi-run) and module ablation studies were conducted using only the best-performing LLM identified overall, any bias or variation in LLM-based evaluation does not influence the conclusions in those sections.

In summary, we thought that, beyond combining multi-judge and expert scoring to reduce bias, more detailed evaluation criteria (as specified in the prompt template) and a greater proportion of non-LLM-based metrics in the overall score can further mitigate LLM bias. This ensures that LLMs with genuinely strong overall performance are consistently ranked at the top.

## Text S2. Execution-grounded checks for the agent-generated codes

Notably, in addition to evaluating code with LLMs and standard metrics/tools (AST and ROUGE), we incorporated a lightweight execution-grounded check from two aspects: data preprocessing outputs and two manually defined checks.

For data preprocessing outputs, we extracted the data preprocessing code snippet, executed it, and verified the number of cells and genes in the processed data. If the results exactly matched those from the reference code, the implementation was considered consistent. Otherwise, it was marked as inconsistent. The consistency rate during the preprocessing stage is presented in Additional file 1: Fig. S3b.

| Task type | Execution-grounded checks |
|---|---|
| Batch correction | (1) adata.obs[“batch”]<br>(2) Check that adata.obsm[“*corresponding_key*”] is a vector whose length equals the number of cells after preprocessing |
| Cell annotation | (1) adata.obs[“cell_type”]<br>(2) Check whether all annotations come from the predefined vocabulary and whether their count equals the number of cells |
| Dynamic analysis | (1) n_cluster>20<br>(2) The graph has few isolated nodes and its connectivity exceeds 80% |
| Perturbation | (1) adata.obs[“ctrl”]<br>(2) Check if the output is a perturbed differential-expression matrix |
| Spatial deconvolution | (1) np.intersect1d(sc_adata.var_names, st_adata.var_names), intersection genes for mapping<br>(2) The output should have the first dimensions equal to the number of spots |
| Gene imputation | (1) spatial_data.var_names<br>(2) Post-imputation expression values must be non-negative |
| Multiomics analysis | (1) adata.obs.index<br>(2) The joint embedding dimension equals the number of intersection cells |
| HVGs selection | (1) adata.X after PCA<br>(2) The output should be a gene list |

Additional file 1: Fig. S3c. Execution-grounded checks for eight representative task types

For two manually defined checks, we selected 13 tasks (i.e., scanvi, scgen-integration, cellassign, decoupler, paga, cellrank, scgen, contrastivevi, totalvi, destvi, tangram, gimvi, and triku) from 8 representative task categories (i.e., batch correction, cell annotation, dynamic analysis, perturbation, spatial deconvolution, gene imputation, multiomics analysis and HVGs selection) as an evaluation subset (Additional file 1: Fig. S3c). For these checks, we first decompose the code to enable intermediate output validation. Specifically, a LLM (GPT-4o) is leveraged to separate the code according to different intermediate checkpoints. This is followed by further inspection by human experts, ultimately transforming the generated codes by agents and groundtruth scripts into multiple code fragments dedicated to execution-grounded checks. Each fragment is executed independently to obtain the outputs of the variables under inspection, which are then compared against the outputs from the corresponding groundtruth script. The comparison process employed a combination of manual and automated verification. We used a binary 0-1 scoring system. For numerical verification, a score of 1 was assigned if the agent-generated code produced numerical results identical to the ground truth; otherwise, it received a score of 0. For validity judgment verification, a score of 1 was given if the output met our predefined validity criteria; otherwise, the score was 0. Additional file 1: Fig. S3e provided detailed execution-grounded check results for three agent frameworks and eight LLMs. All these results serve as a supplementary evaluation of the code. We found that Grok3-beta, GPT-4.1 and Sonnet-3.7 also emerged as the top performing models across frameworks in the intermediate verification, which aligns with the main conclusion of this study.

## Text S3. Complete titles of the supplementary tables

Table S1. Benchmark Tasks Information

Table S2. Multiple-datasets Information

Table S3. Prompts for 50 single-cell omics analysis tasks

Table S4. Result consistency across agent frameworks and LLMs

Table S5. Evaluation prompts

Table S6. Cross-judge evaluation

Table S7. System Prompts for Ablation experiments on AutoGEN